\documentclass[fleqn,usenatbib,useAMS]{mnras}

\usepackage{graphicx}	% Including figure files
\usepackage{amsmath}	% Advanced maths commands
\usepackage{amssymb}	% Extra maths symbols
\usepackage{multicol}        % Multi-column entries in tables
\usepackage{bm}		% Bold maths symbols, including upright Greek
\usepackage{pdflscape}	% Landscape pages
\usepackage{xcolor}	% more colors

\usepackage[T1]{fontenc}
\usepackage{ae,aecompl}

\usepackage{newtxtext,newtxmath}

\newcommand{\mdot}{\ensuremath{{\rm M}_\odot\,{\rm yr}^{-1}}}

\title[(Sub)stellar winds using white dwarfs]{Measurement of stellar and substellar winds using white dwarf hosts}

\author[N. Walters et al.]{N. Walters,$^1$\thanks{E-mail: nikolay.walters.15@ucl.ac.uk}
J. Farihi,$^1$
P. Dufour,$^2$ 
J. S. Pineda,$^3$
and R. G. Izzard$^4$
\medskip
\\
$^1$Department of Physics and Astronomy, University College London, London WC1E 6BT, UK\\
$^2$D\'epartement de Physique, Universit\'e de Montr\'eal, Montr\'eal, Qu\'ebec H3C 3J7, Canada\\
$^3$Laboratory for Atmospheric and Space Physics, University of Colorado Boulder, Boulder CO, 80303, USA\\
$^4$Astrophysics Group, Department of Physics, University of Surrey, Guildford GU2 7XH, UK}
\date{Accepted XXX. Received YYY; in original form ZZZ}
\begin{document}

%\label{firstpage}

\maketitle

\begin{abstract}

White dwarfs stars are known to be polluted by their active planetary systems, but little attention has been paid to the accretion of wind from low-mass companions.  The capture of stellar or substellar wind by white dwarfs is one of few methods available to astronomers which can assess mass-loss rates from unevolved stars and brown dwarfs, and the only known method to extract their chemical compositions.  In this work, four white dwarfs with closely-orbiting, L-type brown dwarf companions are studied to place limits on the accretion of a substellar wind, with one case of a detection, and at an extremely non-solar abundance $m_{\rm Na}/m_{\rm Ca}>900$.  The mass-loss rates and upper limits are tied to accretion in the white dwarfs, based on limiting cases for how the wind is captured, and compared with known cases of wind pollution from close M dwarf companions, which manifest in solar proportions between all elements detected.  For wind captured in a Bondi-Hoyle flow, mass-loss limits $\dot M\lesssim 5\times10^{-17}$\,\mdot are established for three L dwarfs, while for M dwarfs polluting their hosts, winds in the range $10^{-13} - 10^{-16}$\,\mdot are found.  The latter compares well with the $\dot M\sim 10^{-13} - 10^{-15}$\,\mdot estimates obtained for nearby, isolated M dwarfs using Ly$\upalpha$ to probe their astropsheres.  These results demonstrate that white dwarfs are highly-sensitive stellar and substellar wind detectors, where further work on the actual captured wind flow is needed.

\end{abstract}


\section{Introduction}

Stellar wind can be defined as the flow of plasma or gas that is ejected from the upper atmosphere of stars.  Such released material plays a crucial role in enriching the interstellar medium which in turn will be assimilated into the next generation of stars and planets. Despite extensive studies of the solar wind (e.g.,\ \citealt{2021Vidotto}) and asymptotic giant branch (AGB) stars that have an exceptional mass-loss rate due to stellar wind \citep{2018Hofner}, relatively little is known about the mechanism, composition, and mass-loss rates in cool, sub-solar and low-mass stars. 

The most common stars in the Milky Way are M dwarfs, numbering approximately 60 billion in total \citep{2016vanVledder}, where a handful of the nearest and brightest have mass-loss estimates derived from astrospheric Ly$\upalpha$ absorption \citep{2002Wood, 2005Wood, 2021Wood}.  While M dwarf winds are virtually negligible compared to those of massive stars, their mass loss plays a vital role in the context of exoplanet habitability \citep{2007Khodachenko, 2020Mesquita, 2023Ridgway}, and is a key observable for models of their interior, rotation, and magnetic field evolution \citep{2012Reiners, 2014Vidotto}. 

At the bottom of the main sequence and across the substellar boundary, there is little empirical information on intrinsic mass loss from ultracool dwarfs.  It is well known that brown dwarfs tend to be fast rotators \citep{2003Joergens, 2015Scholz, 2021Tannock}, show signs of activity \citep{2000Hawley, 2003Mohanty, 2008Reiners,2016Schmidt}, and can be strongly magnetic \citep{2009Berger, 2012Route, 2018Kao}.  However, magnetic braking in ultracool dwarfs appears to be inefficient, with observational indications that spin-down increases from $\sim 5$ to over 10\,Gyr within the upper half of the L dwarf sequence \citep{2008Reiners}. These long braking times are indicative of the changes in the entirety of the magnetosphere, suggesting a transition in angular momentum evolution.

A clear decline in X-ray emissions into the L dwarf regimes suggests a lack of coronae, and hence non-existent winds \citep{2017Pineda}.  Coupled with changes in topology \citep{2009Schrijver}, this suggests that ultracool dwarfs have magnetospheric environments distinct from those of warmer stars.  In other words, despite the overlap of observable stellar properties between warmer M dwarfs and ultracool dwarfs, the former likely have coronae and winds, while the latter have such features strongly diminished.  Previous studies (e.g.\ \citealt{Pineda2016}), have proposed that the cutoff point for brown dwarf chromospheres lies around spectral type L5.
 
The photospheres of white dwarfs can act as detectors for any heavy elements accreted from their circumstellar environments, including wind from close stellar or substellar companions.  Owing to the high surface gravities in these compact stars, any primordial heavy elements can only be sustained for a brief period in the photosphere, and only to a certain degree while $T_{\rm eff} \gtrsim 25\,000$\,K \citep{1995Chayer, 2014Barstow}. At cooler temperatures, any metals rapidly sink from white dwarf atmospheres on timescales that are always short compared to their evolutionary ages \citep{1979Fontaine, 1979Vauclair}. The diffusion of metals in white dwarf atmospheres is well understood, as is the strict requirement for an external source, and thus the presence of any heavy elements can be interpreted directly as ongoing accretion \citep{1992Dupuis,1993aDupuis,1993bDupuis}.

In pioneering work on the nature of white dwarf pollution, \citet{2003Zuckerman} found that photospheric metals were over-represented in a sample of ten, spatially-unresolved binaries with M dwarf companions; 60 per cent versus 25 per cent for isolated white dwarfs.  Because M dwarfs should undergo mass loss via stellar wind, and those in potentially close pairs appeared to often have a polluted white dwarf host, it was postulated that mass transfer is occurring via stellar wind, and that the mass-loss rate could in principle be estimated.  Based on this scenario, M dwarf winds have been estimated using white dwarf hosts \citep{2006Debes}, although in some cases with highly uncertain orbital separations \citep{2010Farihi_a}.  Nevertheless, the capture of stellar -- and possibly substellar -- wind by white dwarfs remains one of only two known methods to measure intrinsic mass-loss rates in unevolved stars. 

This paper analyzes four known white dwarfs with brown dwarf companions; specifically GD\,1400, WD\,0137--349, NLTT\,5306, and SDSS\,J141126.20+200911.1 (hereafter SDSS\,J1411). The study performs a sensitive search for atmospheric metals using deep co-adds of archival spectra, which in turn constrains any mass loss of the substellar companions through intrinsic winds.  The data and methodology are outlined in Section~\ref{sec:obs}, while Section~\ref{sec:da} describes the framework in which the companion mass loss is calculated.  Section~\ref{sec:res} presents the results, with evidence of wind capture found in NLTT\,5306, and upper limits provided for the three others.  The discussion in Section~\ref{sec:dis} compares the mass-loss rates for M and L dwarfs, contextualises it with prior work on stellar wind determinations, and finishes with an examination of the mass-losing L dwarf NLTT\,5306B.

\section{Observations and Data}\label{sec:obs}

Archival data from 8-m class observatories were searched for optical spectra of known white dwarf + brown dwarf, short-period binaries, with useful data described below in detail. The aim was to construct a sample of close substellar companions that might pollute their white dwarf hosts with any intrinsic mass loss (wind) in the same manner as occurs for close M dwarf companions to white dwarfs \citep{2003Zuckerman}.  For this reason, the search was limited to white dwarfs with $T_{\rm eff} \lesssim 25\,000$\,K, so that any photospheric heavy elements could be confidently attributed to ongoing pollution (as opposed to selective radiation pressure; \citealt{2006Koester}).  It was also required that individual spectra have signal-to-noise (S/N) $>5$ in the region of H$\upalpha$ and H$\upbeta$ so these lines could be confidently fitted and shifted to the white dwarf rest frame for co-adding spectra. 

These selection criteria resulted in four targets for which archival data were extracted and analysed\footnote{The binary SDSS\,J155720.77+091624.6 has been shown to be highly polluted by a circumbinary debris disk and thus cannot be used to search for substellar companion wind accretion.}:  GD\,1400 \citep{2022Walters}, WD\,0137--349 \citep{2006Maxted}, NLTT\,5306 \citep{2013Steele}, and SDSS\,J1411 \citep{2013Beuermann,2014Littlefair}.  Previous work indicates these targets are substellar survivors of at least one common envelope stage, and their white dwarf hosts are susceptible to wind accretion established on the orbital and stellar parameters.  All archival data were reduced from raw files, even in the cases where reduced data products were available.  While the focus was initially to look for the Ca\,{\sc ii} K line at 3934\,\AA, it was necessary to correct for the white dwarf radial velocity using the H$\upalpha$ line cores, and thus the full optical range was extracted for all available data sets.

\subsection{GD\,1400}

A total of 34 spectra were available for GD\,1400 that were obtained with the European Southern Observatory (ESO) Very Large Telescope (VLT) Ultraviolet and Visual Echelle Spectrograph (UVES; \citealt{2000Dekker}). The spectra cover H$\upalpha$ and higher Balmer lines at resolving power $R\approx18\,500$ \citep{2001Napiwotzki}, with the vast majority of exposures acquired under programme 077.D-0673 (PI: Burleigh) in 2006 July. All data were extracted in a semi-automatic way using the ESO Reflex Environment ({\sc esoreflex}; \citealt{2013Freudling}) UVES pipeline version 6.1.6 \citep{2000Ballester}. The standard recipes were used to optimally extract and wavelength calibrate each spectrum. The average S/N was estimated to be 30 in the H$\upalpha$ and H$\upbeta$ regions.

\subsection{WD\,0137--349}

For this binary, 70 archival optical spectra acquired with UVES were analysed. Originally, these spectra were obtained as part of programme 276.D-5014 (PI: Maxted) and 079.C-0683 (PI: Burleigh) between 2005 and 2007. The spectral resolving power varies across arms and setups but is usually between 20\,000 and 30\,000. All data were retrieved and reduced similarly to GD\,1400, with a representative S/N of 15 per spectrum. Additional medium-resolution optical spectra were also downloaded from the ESO archive, for observations made with X-shooter on the VLT \citep{2011Vernet}, under program 093.C-0211 (PI: Casewell) in 2014 August.  The data were reduced using the {\sc esoreflex} X-shooter pipeline v3.5.3 \citep {2010Modigliani} in STARE mode to avoid the automatic co-addition of multiple frames. The estimated S/N is 30 in the VIS arm and 75 in the UVB arm, per average exposure.

\subsection{NLTT\,5306}

There are 24 X-shooter spectra of this target in the ESO archive, which were taken in the NODDING observing mode but reduced in the STARE mode for the same reason as discussed above. These spectra were acquired under two programmes, 085.D-0144 (PI: Steele; \citealt{2013Steele}) and 093.C-0211 (PI: Casewell; \citealt{2019Longstaff}), where the first obtained four exposures on 2010 September 5, and the second acquired the remaining 20 spectra on 2014 August 30.  For this work, both the UVB and VIS arms were reduced, and cover Balmer lines with spectral resolving power 5400 and 8900, respectively. The average S/N is estimated to be 20 in the UVB arm and 17 in the VIS arm, per typical exposure. Further details on these observations can be found in the literature \citep{2013Steele,2019Longstaff}.

\subsection{SDSS\,J141126.20$+$200911.1}

A total of 37 UVB and 28 VIS X-shooter spectra were retrieved for this white dwarf + brown dwarf system.  The data were acquired on 2014 April 19--20 under ESO programme 192.D-0270 (PI: Parsons; \citealt{2014Littlefair}). The UVB exposure time was 450\,s, whereas a longer integration of 600\,s was used in the VIS arm, resulting in fewer of these spectra. The resolving power is approximately 5400 in the UVB and 8900 in the VIS arm. The S/N averages around 12 across the UVB spectra and 10 for the VIS data.

\section{Data Analysis}\label{sec:da}

\subsection{Fitting Balmer features}

To obtain radial velocity information from each spectrum, Balmer lines were fitted using a non-linear, least-square minimisation algorithm from the \textsc{lmfit} python package \citep{2016Newville}. The fitting function for every absorption feature was a Gaussian, or a combination of a Gaussian and a Lorentzian profile if the sharp non-local thermodynamic equilibrium core could be distinguished. Where possible, up to three Balmer features (H$\upalpha$, H$\upbeta$, and H$\upgamma$) were fitted simultaneously by keeping the velocity parameter shared across the fits. Once velocities were obtained for each target, a barycentric correction was applied using the \textsc{astropy} package, where resulting values were fitted with a sine curve. The radial velocity amplitude and relative phase were determined using the residual minimisation routine, but the period was manually set to the orbital period from the literature. The fitted sine model was then used to make relative phase predictions and velocity offsets for individual spectra. A sine fit was deemed sufficient as the orbits should have circularised during the common envelope phase.

\subsection{Construction of white dwarf rest-frame co-added spectra}

Owing to the close proximity of the binary components studied here, illumination by the white dwarf can lead to significant temperature differences between the irradiated (day) and non-irradiated (night) hemispheres of any low-mass companion.  Examples include the brown dwarf companions of interest in this study, WD\,0137--349B \citep{2006Burleigh}, SDSS\,J1411B \citep{2018Casewell_a}, and possibly NLTT\,5306B, but to a much weaker degree \citep{2013Steele, 2023Amaro}.  Of particular relevance to the objectives of this study, the day-side of M and L dwarf companions can be heated into line emission at precisely the same transitions that indicate metal accretion. Thus, an irradiated day-side can interfere with any search for wind pollution, even precluding abundance determinations or upper limits at wavelengths as short as Ca\,{\sc ii} K (\citealt{2003Zuckerman}; examples therein).  Numerous emission lines are known to vary from the day- to night-side of the L dwarf WD\,0137-349B, possibly including Ca\,{\sc ii} K \citep{2017Longstaff}.  Intrinsic brown dwarf emission is also plausible, but unlikely for the older, late L dwarfs examined here \citep{2015Schmidt}.

In order to minimise or eliminate the effects of companion emission lines resulting from irradiation, only those spectra corresponding to night-side phases were utilised. As described above, after obtaining a radial velocity model, each spectrum was velocity shifted to place it into a white dwarf rest frame.  The resulting spectra were then re-sampled with a consistent wavelength range and grid spacing. Next, the S/N ratios of individual exposures were estimated and used as weights to co-add all spectra for each star. Only spectra corresponding to orbital phases 0.3--0.7 were co-added, where phase 0.5 corresponds to the total or maximum obscuration of the irradiated side. This orbital phase range was judged appropriate by inspecting the H$\upalpha$ emission in the highly irradiated companion to WD\,0137--349 as a function of the orbital phase, and is a compromise between (little to no) emission line strength and the S/N of the phase-combined spectra. It is also important to avoid any H$\upalpha$ emission velocities that may overlap with the white dwarf absorption velocity at phases 0.9--0.1. The phases from 0.3 to 0.7 provide an uncontaminated (emission line-free) wavelength region, within which the white dwarf velocity can be readily obtained by absorption line fitting.

An alternative method of creating rest-frame co-adds, that are free of contamination by emission lines was considered. Combining spectra with phases near maxima and minima of radial velocities would, in principle, shift any (day-side) emission lines beyond $\pm100$\,km\,s$^{-1}$ of the white dwarf velocity. This method was applied to the UVES spectra of WD\,0137--349, but was found to offer no significant improvement in terms of S/N or range of wavelengths unaffected by emission lines. Therefore, the night-side-only method was applied throughout, for all four white dwarf + brown dwarf binaries.

These co-added, night-side, rest frame spectra achieved three goals: 1) a region with a minimised day-side emission contribution from the companion, 2) a rest frame in which to search for any weak absorption lines at the same velocity as the white dwarf, and 3) a co-added spectrum with a S/N substantially greater than any individual exposure, in line with $\sqrt{N}$ improvement expectations. Note that because of the phase range requirement of 0.3--0.7 only $\approx40$ per cent of the total available spectra were used in the night-side co-adds. Regions of interest of these co-adds are shown in Figure\,\ref{fig:Fig1}.

The orbital phases were determined using the periods from the following publications: NLTT\,5306 \citep{2013Steele}, GD\,1400 \citep{2022Walters}, WD\,0137--349 \citep{2017Longstaff} and SDSS\,J1411 \citep{2014Littlefair} with the periods given in Table\,\ref{tab:tab1}. These values were verified by identifying the most significant photometric period from the {\em Transiting Exoplanet Survey Satellite} ({\em TESS}; \citealt{2015Ricker}) light curves. This was not possible for GD\,1400 owing to a lack of significant signal at the expected orbital frequency. However, for the remaining three binaries, the {\em TESS} periods were found to be within 0.02 per cent agreement with the published values.

\subsection{Abundance determinations and upper limits}

These rest-frame co-adds were then used to search for heavy element absorption lines, such as those that might be captured as wind from the substellar companion.  The entire available optical wavelength range was scrutinized for any possible photospheric absorption, with particular attention given to Ca\,{\sc ii}\,K, as this transition is well known to be the most prominent optical line in polluted white dwarfs for a wide range of effective temperatures, and across both hydrogen- and helium-rich atmospheres \citep{2003Zuckerman,2005Koester}. Perhaps surprisingly, while the Na\,{\sc i}\,D doublet is not often seen in polluted white dwarfs \citep{2017Hollands,2019Swan}, the archival X-shooter spectra of NLTT\,5306 exhibits these lines, which are always observed at the velocity of the white dwarf \citep{2019Longstaff}. Because of this unexpected pollution signature, this element and wavelength region also received special attention.

The four white dwarfs with substellar companions were fitted with atmospheric models to obtain $T_{\rm eff}$ and $\log\,g$ following the standard technique of Balmer line fitting \citep{2009Tremblay}. As a precaution in the case that emission from the secondary was present, the H$\upalpha$ region was not used in the fitting procedure, but instead H$\upbeta$ to H8 were fitted. The resulting stellar parameters are listed in Table\,\ref{tab:tab1}. 

To determine abundances or upper limits for both Na and Ca, white dwarf atmospheric models were used (\citealt{2019Coutu} and references therein) to produce a grid of synthetic spectra with the derived physical parameters of the observed stars.  The limits (or determinations) were made by fitting the expected positions (or detections) of the Ca\,{\sc ii}\,K and Na\,{\sc i}\,D lines using a standard methodology \citep{2012Dufour}, where upper limits are established by simulating an absorption line that should have been detected at the photospheric velocity, for the corresponding S/N and spectral resolution.

\subsection{Brown dwarf mass-loss rates}\label{sec:MainMethod}

In this section, Ca is used as a typical example of observed white dwarf pollution, but the same methodology applies to all photospheric metals considered in this work. The accretion rate for any heavy element is calculated by considering its total mass in the outer, fully mixed layer of the star, which corresponds to the convection zone in cases where the star is sufficiently cool.  Here, $M_{\rm cvz}$ is used for convenience, but for stars with radiative or stable atmospheres, it is simply the mass of the atmosphere above a Rosseland optical depth of 5 \citep{2009Koester}. This layer has a characteristic diffusion or sinking timescale for Ca ($\uptau_{\rm Ca}$), which will be within a factor of a few of the sinking times for other heavy elements \citep{2020Koester}. In the case of a steady-state balance of accretion and diffusion, where the sinking timescale is much shorter than the ongoing accumulation of heavy elements, the accretion rate can be expressed as:

\begin{equation}
\dot{M}_{\rm Ca} = \frac{X_{\rm Ca}M_{\rm cvz}}{\uptau_{\rm Ca}}
\end{equation}

\smallskip
\noindent
where $X_{\rm Ca}$ is the mass fraction of Ca atoms relative to the dominant atmospheric species (hydrogen is used here, but it could also be helium).  The quantities $M_{\rm cvz}$ and $\uptau_{\rm Ca}$ come from diffusion models based on an individual set of stellar parameters.  In the case of a hydrogen-dominated atmosphere, the Ca mass fraction is:

\begin{equation}
X_{\rm Ca} = \frac{40.078\,n_{\rm Ca}}{1.008\,n_{\rm H}} = \frac{40.078\times10^{[\rm {Ca/H}]}}{1.008}
\end{equation}

\smallskip
\noindent
where $n_{\rm Ca}$ and $n_{\rm H}$ are the number of calcium and hydrogen atoms, respectively, and [Ca/H] $=\log(n_{\rm Ca}/n_{\rm H})$ is determined directly from the fitting of atmospheric models to the spectroscopic data.

To estimate a total mass accretion rate from a single element, a reasonable assumption about the nature of the material must be made (e.g.\ solar, interstellar medium, rocky, icy), so that its relative abundance can be estimated.  For example, by mass, Ca is 1.6 per cent of the bulk Earth \citep{1995Allegre}, but only $7\times10^{-5}$ of the solar photosphere \citep{2003Lodders}.  In this work, solar proportions are assumed as this should broadly reflect the bulk abundances of stars and brown dwarfs at formation.  The total mass accretion rate is estimated by dividing the steady-state accretion rate for Ca by its overall abundance in the infalling material.  Here, $\dot{M_1}$ refers to this total accretion rate onto the white dwarf primary, while $\dot{M_2}$ refers to the corresponding total mass lost from the secondary via (sub)stellar wind onto $M_1$. 

For captured stellar wind with an $\dot{M}_1$ determination (or limit), a few reasonable assumptions can be made to approximate the mass-loss rate from the companion.  The first is a simple model of spherically-symmetric mass loss, which provides a density via the mass continuity equation:

\begin{equation}
\dot{M}_2 = 4\pi \rho v_2 r^2
\end{equation}

\smallskip
\noindent
where $\rho$ is the mass density, and $v_2$ is the velocity of material at a distance $r$ from $M_2$.  It is assumed here that $v_2$ is the escape speed of the companion and is a constant. This is approximately true for the bulk of the solar wind that emerges near the escape speed. However, for the objects studied here, the terminal velocity would likely be similar to the escape speed but reached at a distance of hundreds of R$_\odot$. If the wind is still accelerating this may lead to an overestimation by a significant factor. The precise wind speed profile is beyond the scope of this work, but future modelling would better constrain the actual mass-loss rates for low-mass stars and brown dwarfs.

The $M_2$ radius for a low-mass star or brown dwarf is non-negligible relative to the semimajor axis in close binaries such as those studied here.  For a spherical wind\footnote{The precise flow structure should be addressed with future modelling.} that originates at the surface of such a companion, only material emanating from a narrow range of solid angles (i.e.\ only near the sub-stellar point) has any chance of being captured by the white dwarf, so that $r \approx a - R_2$, where $a$ is the semimajor axis and $R_2$ is the radius of the companion. The velocity of the flow relative to the white dwarf can be approximated as $v_1^2 = v_2^2 + v_{\rm orb}^2$, where the latter term is the Keplerian speed of the white dwarf.

The remaining quantity to be estimated is the density at the accretion radius $R_A$, where two models are considered here: the ideal case of Bondi-Hoyle-Littleton accretion \citep{1952Bondi}, and simple gravitational capture, also known as Eddington accretion \citep{1926Eddington}.  In each case, the model describes a distinct $R_A$, within which all material is captured; a Bondi-Hoyle flow is an ideal fluid case where there is zero transverse momentum downstream of the accreting source, whereas an Eddington geometry is simply where gravity is sufficient to alter the flow trajectory onto the star. A detailed discussion on the application of these models to white dwarf accretion can be found in \citet{2010Farihi_a}.

Starting with the Bondi-Hoyle accretion, the material density at $R_A$ is given by

\begin{equation}
\rho_{\rm BH} = \frac{\dot{M_1}v_1^3}{4\pi G^2 M_1^2}
\end{equation}

\smallskip
\noindent
where $G$ is the gravitational constant.  By making appropriate substitutions in Equation (3), the following mass-loss estimate is obtained:

\begin{equation}
\dot{M}_2 = \frac{\dot{M}_1 v_1^4 \, r^2}{G^2 M_1^2}
\end{equation}

\smallskip
\noindent
For Eddington accretion, the material density at $R_A$ depends on $R_1$ as:

\begin{equation}
\rho_{\rm Edd} = \frac{\dot{M}_1 v_1}{2\pi GM_1 R_1}
\end{equation}

\smallskip
\noindent
where appropriate substitutions lead to the following mass-loss rate:

\begin{equation}
\dot{M}_2 = \frac{2\dot{M}_1 v_1^2 \, r^2}{GM_1R_1}
\end{equation}

\smallskip
\noindent
Using NLTT\,5306 as an example, $\rho_{\rm Edd}/\rho_{\rm BH} \approx 40$, which is the same ratio between the two corresponding mass-loss estimates.  All resulting mass-loss and accretion rates, as well as upper limits, are provided in three columns of Table\,\ref{tab:tab1}. 

The diffusion timescales and fully-mixed, atmospheric mass fractions were determined using bilinear interpolation of hydrogen-rich model grids with convective overshoot \citep{2020Koester}, where all four stars in the sample have been characterised as hydrogen atmosphere white dwarfs that manifest as spectral type DA. Steady-state accretion is a safe assumption, as the wind accretion timescales are the white dwarf cooling ages ($\sim10^8-10^9$\,yr), while the heavy element sinking timescales are at least $1000\times$ shorter.

Evolutionary models for brown dwarfs \citep{2015Baraffe} were employed to estimate parameters of the potentially mass-losing substellar secondaries, such as $R_2$ and $\log g$.  The age of 2\,Gyr was chosen for this purpose, as brown dwarfs should be effectively fully contracted by then. However, determining the ages of brown dwarfs from their white dwarf companions is not feasible, as the white dwarfs have undergone binary (non-isolated) evolution. In this scenario, the white dwarf cooling ages act as the minimum ages for the systems.

In the case that the system is still young and the radius is not fully contracted there would be a modest increase in the mass-loss rate. For example, a 500\,Myr brown dwarf would have a 10 per cent larger radius than a fully contracted older counterpart \citep{2015Baraffe}.  Such a larger radius would result in a 40 per cent increase in the mass-loss rate in the case of a Bondi-Hoyle flow, or a 60 per cent increase in the case of gravitational capture.

\section{Results}\label{sec:res}

The results section is organized as follows.  First, metal pollution resulting from substellar wind is evaluated in terms of upper limit or determined accretion rates. Second, corresponding mass-loss rates and limits are derived for the four systems with brown dwarf companions detailed in Section~\ref{sec:obs}.  Third, these substellar wind estimates are then compared to that of several M dwarfs known to be polluting their host white dwarfs via wind capture.

\begin{table*}
%\centering
\begin{center}
\caption{Adopted stellar parameters, abundance determinations, accretion and mass-loss rates for (sub)stellar winds captured by white dwarfs\label{tab:tab1}.}
\begin{tabular}{lrlrlrrrrrrl} 

\hline
    &\multicolumn{2}{c}{Adopted}
    &\multicolumn{2}{c}{Estimated}
    &Orbital
    &\multicolumn{2}{c}{Abundances} 
    &\multicolumn{1}{c}{Accretion}
    &\multicolumn{1}{c}{BH Capture}
    &\multicolumn{1}{c}{Edd Capture}
    &\\
    
    Binary          
    &$T_{\rm eff}$  
    &$\log\,g$  
    &$M_2$  
    &SpT        
    &Period
    &\multicolumn{1}{c}{[Ca/H]} 
    &\multicolumn{1}{c}{[Na/H]}             
    &\multicolumn{1}{c}{$\dot{M}_1$}    
    &\multicolumn{1}{c}{$-\dot{M}_2$}
    &\multicolumn{1}{c}{$-\dot{M}_2$}  
    &Refs\\
                
    &(K)         
    &(cgs) 
    &(M$_{\rm Jup}$)
    &   
    &(h)  
    &       
    &                   
    &\multicolumn{1}{c}{(\mdot)}          
    &\multicolumn{1}{c}{(\mdot)}      
    &\multicolumn{1}{c}{(\mdot)}          
    &\\

\hline

{\bf Low-mass stars}:\\

Case\,1         &15\,500    &8.07   &380    &M3 &16.0   &$-8.1$     &...        &$2.6\times10^{-17}$    &$1.9\times10^{-15}$    &$1.0\times10^{-13}$    &1,2,3,4\\

PG\,1026+002    &17\,200    &7.96   &380    &M5 &14.3   &$-8.6$     &...        &$6.0\times10^{-18}$    &$4.4\times10^{-16}$    &$2.0\times10^{-14}$    &1,2,3,5,6\\

LHS\,1660       &7500       &7.70   &190    &M5 &7.3    &$-9.3$     &...        &$3.1\times10^{-16}$    &$9.9\times10^{-15}$    &$3.2\times10^{-13}$    &1,2,7,8\\

PG\,2257+162    &24\,300    &7.51   &190    &M5 &7.7    &...        &...        &$3.1\times10^{-17}$    &$1.3\times10^{-16}$    &$3.2\times10^{-15}$    &1,9,8\\

BPM\,6502       &22\,600    &7.84   &150    &M5 &8.1    &...        &...        &$1.4\times10^{-17}$    &$2.6\times10^{-16}$   &$1.3\times10^{-14}$   &1,3,10\\

LTT\,560        &7500       &7.75   &150    &M6 &3.5    &$-7.5$     &...        &$8.5\times10^{-15}$    &$9.4\times10^{-14}$    &$3.1\times10^{-12}$    &1,11\\

Rubin\,80       &8200       &7.75   &100    &M7 &4.0    &$-8.3$     &...        &$8.7\times10^{-16}$    &$1.0\times10^{-14}$    &$3.7\times10^{-13}$    &1,2,3,6,12\\

GD\,448         &19\,700    &7.49   &100    &M7 &2.5    &...        &...        &$1.2\times10^{-17}$    &$7.0\times10^{-17}$    &$1.9\times10^{-15}$    &1,9,13\\

\hline

{\bf Brown dwarfs}:\\

NLTT\,5306      &7700       &7.49   &60     &L5 &1.7    &$<-11.0$   &$-7.7$     &$3.3\times10^{-15}$    &$5.6\times10^{-15}$    &$2.1\times10^{-13}$    &1,14\\

GD\,1400        &11\,400    &8.17   &70     &L6 &10.0   &$<-9.8$    &$<-7.3$    &$<2.4\times10^{-18}$   &$<5.5\times10^{-17}$   &$<4.3\times10^{-15}$   &1,15\\

WD\,0137--349   &17\,600    &7.58   &60     &L7 &1.9    &$<-7.9$    &$<-5.7$    &$<2.2\times10^{-17}$   &$<5.1\times10^{-17}$   &$<2.0\times10^{-15}$   &1,16,17\\

SDSS\,J1411     &11\,800    &8.01   &50     &L8 &2.0    &$<-8.2$    &$<-6.2$    &$<3.9\times10^{-17}$   &$<4.6\times10^{-17}$   &$<4.1\times10^{-15}$   &1,18,19\\

\hline

\end{tabular}
\end{center}

\flushleft
{\em Notes.}  The Ca abundance limits for GD\,1400 and WD\,0137--349 were determined from night-side UVES spectra, while those for NTT\,5306 and SDSS\,J1411 were obtained using night-side X-shooter data. Stellar parameters and abundances for M dwarf hosts are taken from the literature, where GD\,448 and PG\,2257+162 have only Si and C abundances from ultraviolet observations. Spectral types and masses for the companions were estimated from the literature, and from the absolute $K$-band magnitude in some cases. For the low-mass stars, accretion and mass-loss rates are averages based on all available abundances, including a few not shown here or in the left panel of Figure\,\ref{fig:Fig2}. For the brown dwarfs, the rate limits are founded on the more constraining Ca abundances, with the exception of NLTT\,5306, for which a Na abundance was determined and used to estimate the rates.  All (single-element) accretion rates assume the metal is present in solar proportions.\\
\smallskip
{\em References}: 
(1) This work;
(2) \citet{2003Zuckerman};
(3) \citet{2011Gianninas};
(4) \citet{1982Lanning}; 
(5) \citet{1993Saffer}; 
(6) \citet{2005Farihi_a}; 
(7) \citet{2007Maxted}; 
(8) \citet{2017Parsons_b};
(9) \citet{2014Koester}; 
(10) \citet{2008Kawka};
(11) \citet{2011Tappert_a};
(12) \citet{2019Ashley}; 
(13) \citet{1998Maxted};
(14) \citet{2013Steele};
(15) \citet{2004Farihi}; 
(16) \citet{2006Maxted};
(17) \citet{2017Longstaff};
(18) \citet{2013Beuermann}; 
(19) \citet{2014Littlefair}.
\end{table*}

\subsection{Brown dwarf winds}\label{sec:BDWind}

\begin{figure}
\includegraphics[width=\columnwidth]{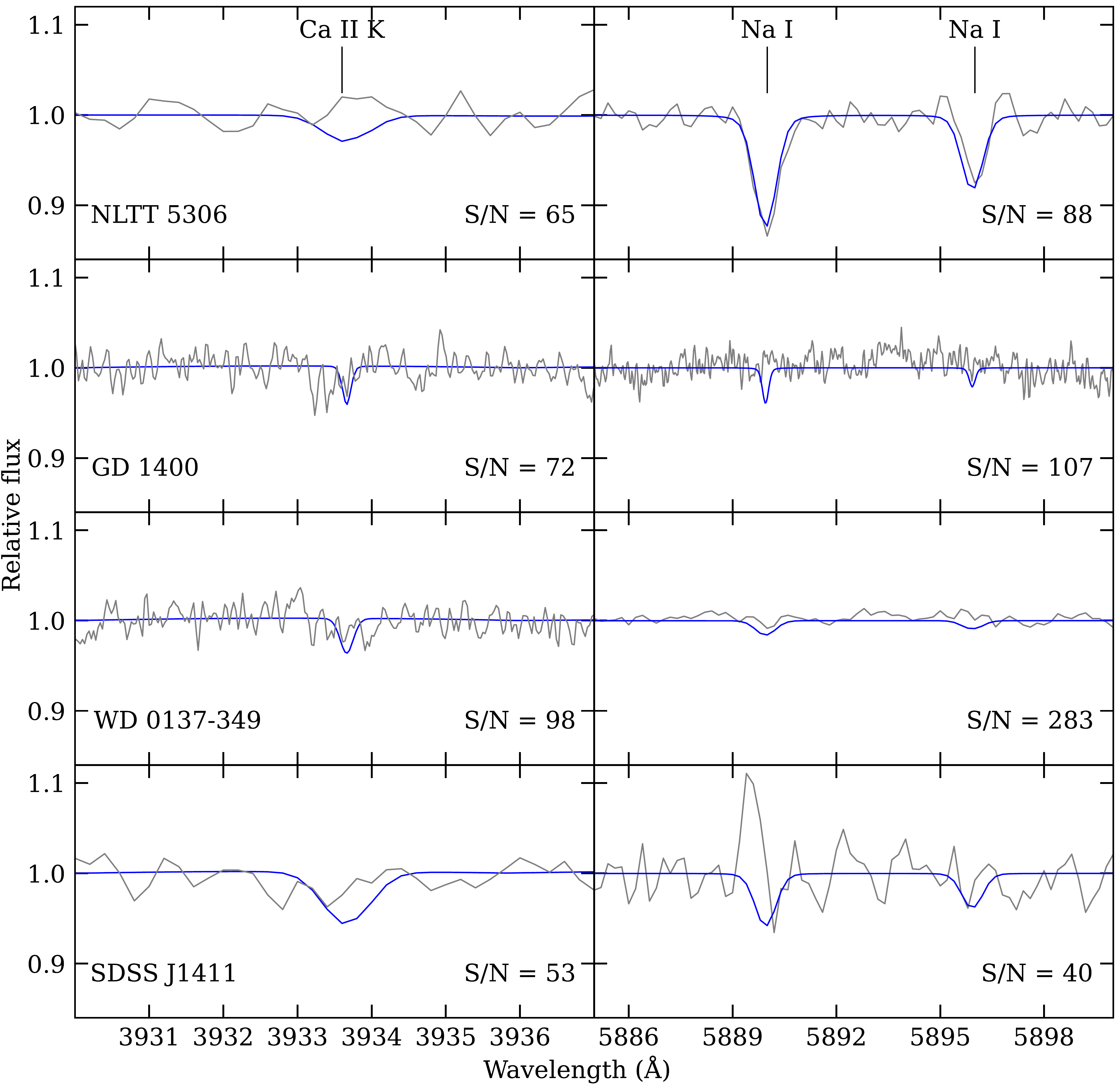}
\vskip 1mm
\caption{The night-side co-added spectra (in grey) of all four white dwarfs hosting L dwarf companions, plotted in the regions of the Ca\,{\sc ii}\,K line (left panels), and Na\,{\sc i} D doublet (right panels), with relevant S/N estimates given in each panel. Atmospheric models (in blue) are overplotted in each case to determine the upper limit abundances of these two elements, with the sole exception of the Na detection in NLTT\,5306.  There are continuum imperfections in some cases (e.g.\ GD\,1400), but these have little effect on the modelling outcomes. All co-added spectra were scrutinised for any additional absorption features, but none were found.}
\label{fig:Fig1}
\end{figure}

It should be emphasized that conventional {\em mass transfer} is not considered here as intrinsic mass loss analogous to the stellar wind. As shown in Table\,\ref{tab:tab1}, a typical estimated upper limit for the accretion onto the white dwarf is $\dot M_1<5\times10^{-17}$\,\mdot.  This is at least three orders of magnitude lower than the solar value of $2\times10^{-14}$\,\mdot \citep{2011Cohen}. For all but NLTT\,5306, there is a lack of detected pollution and thus only corresponding upper limits for any substellar winds. There is a clear Na\,{\sc i} doublet detection in the co-added spectrum of NLTT\,5306, but only weak upper limits for Na in the other three binary systems.  Moreover, there are no detections of Ca (only upper limits) in all four white dwarfs with close substellar companions (Figure\,\ref{fig:Fig1}). Based on these non-detections, the presented accretion and mass capture constraints are calculated from the more constraining Ca limit rather than Na. The only exception is NLTT\,5306, for which the determined Na abundance was used instead of the Ca upper limit.

For additional context, polluted white dwarfs with infrared detections of their circumstellar disks have typical, total inferred mass accretion rates $\dot M\gtrsim10^{-18}$\,\mdot \citep{2009Farihi}, where a common assumption is that Ca is 0.016 of the total accreted mass as it is in the Earth \citep{2016Farihi}.  But while accretion rates for white dwarfs polluted by (sub)stellar winds have been calculated using the same steady-state formalism as done for planetary debris pollution, the limits established here are higher for wind capture, despite the fact that similar abundance sensitivities have been established using the same large telescopes and sensitive spectrographs. Owing to the dominance of hydrogen, Ca has a mass fraction that is roughly $220\times$ smaller in the sun than in the rocky material of the inner solar system, and thus a fixed Ca abundance translates to a total accretion rate that is commensurately higher for solar composition as opposed to terrestrial abundances.

%%%FIGURE 1%%%

Unlike Ca, there is a clear detection of Na in the spectrum of NLTT\,5306, with a determined abundance [Na/H] $= -7.7$. It is important to note that the lack of Na\,{\sc i}\,D detections in GD\,1400, WD\,0137--349, and SDSS\,J1411 are not highly constraining.  For a fixed abundance, the Na\,{\sc i}\,D transition is significantly stronger in $T_{\rm eff}\lesssim10\,000$\,K and cooler white dwarfs \citep{2021Hollands,2022Hollands}.  Thus, for these three warmer white dwarfs with substellar companions, the high S/N spectroscopy is insensitive to the Na abundance detected in NLTT\,5306.  Because the Na abundance upper limits are two orders of magnitude less stringent than those placed on Ca, the latter element provides more sensitive $\dot{M_1}$ and $\dot{M_2}$ limits in accordance with a solar composition wind.

\subsection{Stellar winds}

To provide independent main-sequence mass-loss benchmarks, the methodology described in the previous section is also applied to several known white dwarfs that are polluted by the wind of their M dwarf companions, and where metal abundances are available in the literature. For four of these sources (Case\,1, PG\,1026+002, LHS\,1660, and Rubin\,80) Ca and other heavy element abundances were derived from optical spectroscopy, as described in \citet{2003Zuckerman}. In addition, C and Si abundances for two sources (PG\,2257+162 and GD\,448) were determined using ultraviolet data, as reported in \citet{2014Koester}. The adopted physical parameters from the literature are listed in Table\,\ref{tab:tab1}; in cases where multiple sources were available a preference was given to more recent work and studies that account for the binary nature of the system.

There are two additional, close M dwarf companions to white dwarfs in the literature, where pollution from stellar wind has been detected. The close binary BPM\,6502 has {\em FUSE} ultraviolet spectroscopy, from which photospheric abundances have been derived for C, N, Si, and Fe \citep{2008Kawka}.  Based on these abundances, the white dwarf is likely too warm to exhibit metal absorption via optical spectroscopy. Indeed, an examination of archival UVES observations from the SPY survey \citep{2003Napiwotzki} reveals a DA spectrum with no clear metal absorption, but strong emission lines of Ca\,{\sc ii}\,H \& K from the irradiated companion. Remarkably, one of the most metal-rich white dwarfs known is LTT\,560, which is highly polluted by the wind of its M dwarf companion and shows strong H$\upalpha$ emission. There are even photospheric detections of the elements Sc and Co \citep{2011Tappert_a}, where otherwise these two rare elements have only been detected together in extremely polluted white dwarfs with infrared-bright debris disks \citep{2007Zuckerman,2012Dufour}.  

All eight aforementioned white dwarfs that are polluted by close M dwarf companion winds were re-analysed based on their published heavy element abundances, but with state-of-the-art diffusion parameters \citep{2020Koester}\footnote{The average percentage difference between this work and \citet{2006Debes} for convection zone mass ratio and Ca diffusion timescale is 3 and 32\,per cent, respectively.}.  As in Section~\ref{sec:MainMethod}, each detected element was assumed to be present in solar abundance, and in this way each metal species present provides an independent estimate of the total accretion rate.  For the M dwarf secondaries, stellar parameters were obtained from the model grids of \citet{2015Baraffe}, while white dwarf parameters were generally taken from the literature (N.B.\ the \textit{Gaia} eDR3 white dwarf catalogue entries for these targets are likely inaccurate because of the optical red flux of the companion). 

\begin{figure*}
\includegraphics[width=\linewidth]{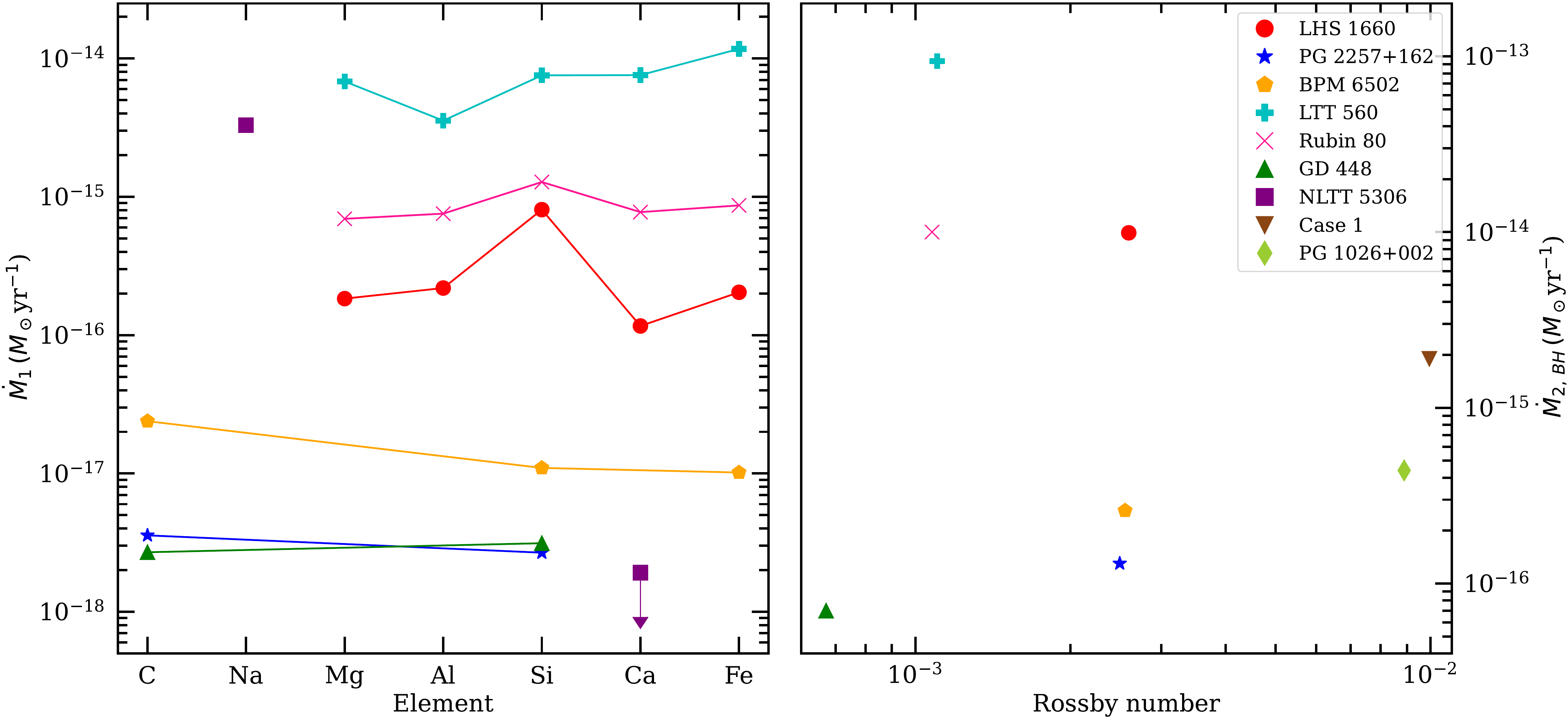}
\vskip 1mm
\caption{On the left are white dwarf accretion rate inferences for all M dwarf hosts in Table\,\ref{tab:tab1} with multiple heavy element abundances, where all elements are assumed to be present in solar ratios.  Case\,1 and PG\,1026$+$002 are not plotted since only the Ca abundance was available.  For the L-dwarf host NLTT\,5306, the ratio of Na accretion rate and Ca upper limit provide a stark contrast to the M dwarf wind accreting stars.  On the right are shown Rossby numbers for the M dwarfs against their mass-loss rates estimated using the Bondi-Hoyle formalism.}
\label{fig:Fig2}
\end{figure*}

The resulting accretion and mass-loss estimates are summarised in Table\,\ref{tab:tab1}, where an average across all detected elements is given in the ninth column (with upper limits established on Ca for L dwarf hosts).  The inferred accretion and mass-loss rates for M dwarfs and the L dwarf and NLTT\,5306 are plotted together in Figure\,\ref{fig:Fig2}, where for each element, $\dot M_1$ is calculated assuming it is present in solar proportions.  With a few minor exceptions, the left-hand panel demonstrates that distinct metal species yield broadly similar estimates for $\dot M_1$ when solar abundances are assumed, thus indicating the captured M dwarf winds are approximately solar in composition.

\section{Discussion}\label{sec:dis}

%%%FIGURE 2%%%

\subsection{Prior work and measurements}

The first work that attempted to estimate M dwarf mass-loss rates using polluted white dwarfs relied on existing Ca abundances in suspected close binary systems and assumed that the wind is captured in a Bondi-Hoyle flow \citep{2006Debes}. Of the six binaries in that study, there are four here in common: Case\,1, PG\,1026+002, LHS\,1660, and Rubin\,80. 

While there is an order-of-magnitude agreement between the prior estimates and those made here using the Bondi-Hoyle prescription in two cases, the other two are notable exceptions. First, the binary orbit of Rubin\,80 was only recently characterized \citep{2019Ashley}, and thus the prior mass-loss estimate suffered from a poorly constrained orbital separation.  Second, the stellar parameters of Case\,1 differ significantly between the new and prior work, with the sinking timescale used here $40\times$ shorter and implying a commensurately higher white dwarf accretion rate, and thus mass-loss rate in the M dwarf.  Third, it should be noted that the remaining two systems in \citet{2006Debes} are now known to be wide binaries \citep{2010Farihi_b}, and thus wind cannot be responsible for any atmospheric pollution (which is unconfirmed for PG\,1210+464, but clearly detected in PG\,1049+103; \citealt{2003Zuckerman}).

According to some studies, M dwarf stellar wind measurements can be obtained from low accretion rate polars (magnetic cataclysmic variables; \citealt{2002Schwope}).  In this scenario, the donor star is underfilling its Roche lobe, and thus mass transfer would only be possible via wind accretion \citep{2009Schwope}. However, in contrast to the M dwarf companions studied here, a strongly magnetic white dwarf may couple to the magnetic field lines of the secondary, and thus have potential to influence stellar wind, which is itself an intrinsically magnetic process.  Above a critical magnetic field strength $B\sim60$\,MG, a white dwarf is capable of collecting the global mass loss from a donor star out to at least several R$_\odot$ \citep{1994Li,2002Webbink,2005Webbink}.  In low accretion rate polars with M dwarf secondaries, the total accretion rates range from around $5\times10^{-14}$ to $6\times10^{-13}$\,\mdot \citep{2005Schmidt,2007Schmidt_b,2010Kafka,2021Parsons}.  

These accretion rate inferences are $2-4$ orders of magnitude larger than the M dwarf wind capture rates determined here in Table\,\ref{tab:tab1}, and are therefore unlikely to represent intrinsic, unenhanced mass loss from low-mass stars.  Such high corresponding mass-loss rates have been questioned previously, especially in similar systems where the donor is substellar, and where they notably exceed the solar mass-loss rate (e.g.\ \citealt{2008Farihi,2017Stelzer}).  Although the surface magnetic field and rotational rates tend to be higher in M dwarfs, the general trend is that activity decreases towards lower stellar temperatures and masses \citep{2011Cranmer,2023Kumar}.  Therefore, low accretion rate polars are not representative of unassisted mass loss, but rather mass transfer that is modulated in the presence of strong white dwarf magnetism.  However, these mass transfer rates provide a useful, high benchmark to which the estimates calculated here for completely detached, non-magnetic binaries can be compared.

The only other successful mass-loss estimates for M dwarfs are made using astrospheric signatures of interactions between their stellar winds and the interstellar medium near their respective astropauses \citep{2005Wood}. The confluence of wind and interstellar medium can produce detectable Ly$\upalpha$ absorption, which can be modelled to constrain the mass-loss rate, where estimates span nearly three orders of magnitude $0.1-60\times10^{-14}$\,\mdot \citep{2021Wood}.  It is noteworthy that the solar wind value lies in the middle of this range, and that winds an order of magnitude stronger are implied by this methodology. In contrast to the M dwarfs studied here, astrospheric observations target nearby, bright, and well-studied stars with constraints on X-ray luminosity, and some indications of magnetic field strength as well as topology.

\subsection{New M dwarf wind estimates}

To interpret the results of this work, it must first be asked whether a Bondi-Hoyle flow or gravitational capture is more realistic, or perhaps another accretion geometry (e.g.\ influenced by white dwarf magnetism). Based on the previously discussed examples of accretion in low-state polars, it is well-known that strong magnetic fields can redirect mass transfer.  Magnetic fields can do work on an incoming flow, and effectively remove transverse momentum downstream, as well as capture material from a wider radius.  While none of the white dwarfs in this study are known to be magnetic, that does not rule out weak magnetism on the order of a few kG or lower \citep{2021Bagnulo}, and which may play a role in assisting the capture of stellar wind. 

Comparing the newly-obtained M dwarf wind estimates with the solar value, the Bondi-Hoyle estimates are all lower or similar to the solar wind mass-loss rate. In contrast, the pure gravitational capture model results in many mass-loss rates that exceed the solar value, and thus suggest that a Bondi-Hoyle flow is the more realistic model. An important caveat is that the close M dwarf companions to white dwarfs are all tidally locked, thus rotating rapidly and highly active.  Instead, comparing the Table\,\ref{tab:tab1} mass-loss estimates with those of the most active M dwarfs in \citet{2021Wood}, gravitational capture estimates appear to be a closer match. Further investigation into the flow geometry, especially the influence of weak white dwarf magnetism, is beyond the scope of this work but would better constrain stellar wind measurements using atmospheric pollution.

There are no indications that the white dwarfs studied here have any influence on the winds of their companion stars, irrespective of the mechanism of wind capture.  However, there is a sensitivity bias, where photospheric heavy elements in white dwarfs are far easier to detect in the ultraviolet than in the optical (e.g.\ \citealt{2014Koester}); the three lowest mass-loss estimates in Table\,\ref{tab:tab1} and Figure\,\ref{fig:Fig2} are the result of ultraviolet observations.  Overall, the rates of mass loss estimated from white dwarf pollution are not significantly different from those measured for isolated M dwarf stars using astrospheric Ly$\upalpha$ absorption \citep{2021Wood}.  However, the M dwarf companions studied here are all tidally locked and have rotation periods of hours, and thus are somewhat distinct from the isolated M dwarfs studied by \citet{2021Wood}.

\subsection{Mass loss and activity across the substellar boundary}

Limits on substellar winds are a new constraint on the activity and structural changes that occur as mass and temperature decrease along the ultracool dwarf sequence, from low-mass stars to brown dwarfs.  While instances of high stellar activity are commonly linked to fast rotational velocities \citep{1972Skumanich,2011Wright}, and spectroscopic observations have shown that many brown dwarfs are indeed fast rotators (e.g.\ \citealt{2003Mohanty, 2010Blake}) -- in some cases rotating at $\approx30$ per cent of their break-up speed \citep{2012Konopacky} -- ultracool dwarfs of spectral type M9 and later exhibit low activity levels \citep{2000Gizis,2003Mohanty,2008Reiners}. This deterioration in the rotation-activity relationship is further supported by X-ray observations of late M-type stars \citep{2014Williams, 2014Cook}. This may be an empirical indication that, in general, L-type brown dwarfs do not generate substantial winds.

This observed change in the rotation-activity relation at the M/L transition could be due to a fundamental change in the dynamo mechanism within ultracool dwarfs, reducing dynamo efficiency and decoupling field generation from rotation rate \citep{2010Reiners}. Another possibility is the growing atmospheric neutrality which weakens the coupling between the ionized atmosphere and magnetic field, consequently rendering magnetic heating ineffective \citep{2012Reiners_b}. As a result, there might be more similarities in the underlying magnetic field generation between brown dwarfs and giant planets \citep{2009Christensen,2011Morin} than in solar-mass stars. In this sense, brown dwarf activity can exhibit itself in the form of an auroral-planet framework rather than conventional stellar winds \citep{2017Pineda}. However, this auroral population is likely distinct from those ultracool dwarfs exhibiting coronal or chromospheric behaviour \citep{2012Stelzer}. Some activity indicators, such as X-rays \citep{2003Tsuboi,2006Stelzer} and H$\upalpha$ emission have been detected in a handful of brown dwarfs \citep{2003Mohanty,2007Schmidt_a,2008Reiners}, implying chromospheric activity, and thus potentially, some mass loss in their outermost layers. 

The right-hand panel of Figure\,\ref{fig:Fig2} is an attempt to contextualize the mass-loss rates of M dwarf companions to white dwarfs, in a manner similar to astrospheric wind estimates, by accounting for the rotation rate (assumed to be identical to the orbital period owing to tidal locking). The figure displays the inferred mass-loss rates for M dwarf companions to white dwarfs, in the Bondi-Hoyle approximation, as a function of the Rossby number (the ratio of the spin period to the convective turnover timescale, following \citealt{2018Wright}). In this diagram, there is a distinct lack of correlation between mass-loss rates and the Rossby number, which should be a reliable indicator of magnetic field strength and stellar activity \citep{2022Reiners}. 

A dependence between mass-loss rates and Rossby number would only be expected in an unsaturated regime of stellar activity.  However, almost all of the M dwarf companions studied here are in the coronally-saturated regime, where stellar winds no longer correlate with rotation \citep{2017Johnstone}, similar to magnetic field saturation and other activity indicators \citep{2022Reiners}. One potential exception is Case\,1 which would require a rotational period of $\sim 5$\,h or less to be fully saturated \citep{2014Reiners}. Further comparison is not possible without detailed information on the properties of the M dwarf companion magnetic fields.

Comparing the results obtained here with those for nearby, bright, and active M dwarfs \citep{2021Wood}, the astrospheric technique has a typical sensitivity to $\dot M\gtrsim10^{-15}$\,\mdot, and identifies mass-loss rates centred near $10^{-14}$\,\mdot, comparable to the solar value.  On the one hand, these astrospheric wind determinations are broadly consistent with the Table\,\ref{tab:tab1} inferences for Eddington capture.  On the other hand, roughly half of the Table\,\ref{tab:tab1} wind estimates made using a Bondi-Hoyle flow are notably near or below the sensitivity for astrospheric wind detection.  It is therefore possible that those M dwarfs with detected astropheres are outliers within a distribution that, for the most part, are more sedate and similar to the Table\,\ref{tab:tab1} results for the Bondi-Hoyle wind capture model.  However, it is important to note that the nearby M dwarfs from \citet{2021Wood} exhibit significantly lower and more typical rotational rates, with the shortest period just under 3\,d.  This complicates a direct comparison since the two samples are likely to be operating in different activity regimes. Finally, the upper limit of $5\times10^{-17}$\,\mdot found for the L-dwarfs using the Bondi-Hoyle formalism, when translated into a mass loss per unit surface area, is comparable to the most sensitive measurements presented in \citet{2021Wood}.  In other words, the white dwarf pollution technique has a similar sensitivity to the more instrumentally demanding Ly$\upalpha$ monitoring that is generally limited to within 7\,pc (due to being outside of cooler, partially neutral material within the Local Bubble).  It should also be noted, that the astrospheric technique inherits systematic uncertainties associated with the methodology that could skew this comparison.

The binary orbital motion may play an underestimated role in determining the actual geometric structure of the wind outflow.  This is plausible, as the bulk of unevolved, low-mass companions have orbital velocities that are the same order of magnitude as their escape velocities.  However, if the wind capture were strongly dependent on orbital velocity, then a correlation between accretion rate and binary period might be seen, yet no such trend is evident. Because the binary mass ratios are not greatly dissimilar (all should have $q=M_2/M_1 \lesssim 0.6$; \citealt{2013Pecaut}), these are unlikely to influence the flow and wind capture mechanism between individual systems. Such considerations underscore the benefit of further work on the wind flow geometry toward white dwarfs from their low-mass stellar and substellar companions, and which will result in more accurate comparisons to stellar wind estimates obtained for isolated stars.

\subsection{The detection of Na in NLTT\,5306}

The translation of the photospheric Na in NLTT\,5306 into a mass-loss rate from an L dwarf is a first.  Because the measurement derives from detectable white dwarf pollution, it is perhaps unsurprising that the nominal mass-loss rate for the brown dwarf is comparable to those estimated for M dwarf companions, under a model that each detected element is present in solar abundance.

In the case of NLTT\,5306, however, the detected Na and upper limit Ca abundances cannot be reconciled with material that is ejected and accreted in solar elemental ratios. This is made strikingly apparent in the left panel of Figure\,\ref{fig:Fig2}, where there is more than three orders of magnitude disparity between the two accretion rates as inferred for solar composition. In contrast, all other wind-polluted white dwarfs appear to be accreting matter with element-to-element ratios broadly consistent with solar abundance material, insofar as the available data allow. 

The steady-state accretion rate for Na and upper limit for Ca yield $m_{\rm Na}/m_{\rm Ca}>900$ for the matter falling onto NLTT\,5306.  This highly unusual mass ratio can be directly compared to the solar and chondritic values where both are $m_{\rm Na}/m_{\rm Ca}\approx0.5$ \citep{2003Lodders}, and {\em three orders of magnitude smaller}. Therefore, the material being accreted by the white dwarf cannot be the result of conventional mass transfer via Roche lobe overflow, as gravity would not differentiate between neutral atoms, molecules, or dust grains.  Moreover, even if Ca is locked up in dust it will be accreted, as radiation pressure on small grains is negligible at the luminosity of most white dwarfs.  Lastly, while the metallicity of many brown dwarfs may be subsolar, their bulk compositions are expected to reflect formation by gravitational collapse and thus be of broadly solar ratios.

Therefore, the pollution in this white dwarf is by a weak substellar wind, and where the observed, intrinsic H$\upalpha$ emission is a hallmark of accretion similar to that observed in other white dwarfs capturing M dwarf winds \citep{1998Maxted,2011Tappert_b, 2011Tappert_a,2013Ribeiro}.  In fact, this telltale sign is prominent in only two cases (NLTT\,5306 and LTT\,560) and detectable but at weaker strength in a third system (RR\,Cae = LHS\,1660); all of which are relatively low-luminosity white dwarfs with $T_{\rm eff}<8000$\,K. These modest signatures of accretion luminosity via H$\upalpha$ emission may be difficult or impossible to detect against the higher intrinsic brightness of white dwarfs that are warmer, and against the intrinsic or irradiatively-driven, chromospheric emission lines of the secondary (precisely the case for GD\,448; \citealt{1998Maxted}). 

The extreme Na/Ca ratio cannot be planetary debris of any known origin, nor can it be material remaining from previous stages of stellar evolution. Post-AGB stars in binaries can have metal-poor surfaces and non-solar abundances because they accrete material from a circumbinary disc \citep{1992Waters, 2019Oomen, 2020Oomen}, where a well-known example is the Red Rectangle \citep{1996Waelkens}.  In this scenario, the accreted material is reduced in refractory elements owing to the formation of dust grains that are expelled by radiation pressure.  Nonetheless, even in those systems $m_{\rm Na}/m_{\rm Ca}\ll900$, and typically enhanced only by a factor of $10-20$ \citep{1992vanWinckel,1998vanWinckel}. Owing to the sinking timescale of heavy elements in white dwarfs, these short-lived post-AGB system signatures cannot persist to the present epoch, over the cooling age of NLTT\,5306 \citep{2001Zijlstra,2013Steele, 2023Amaro}.

Some elemental fractionation is observed in the solar wind and corona, as well as M dwarf coronae. In the case of the sun, low first ionization potential (FIP) elements, such as sodium, can be typically enhanced in abundance by a factor of two to five \citep{2015Laming}, where more pronounced fractionation can be observed locally during solar flares \citep{2015Doschek}.  In contrast, in M dwarfs with $T_{\rm eff}<4000$\,K, low-FIP elements are coronally depleted by a factor of three to four \citep{2018Wood}.  Although such changes in coronal abundances, relative to solar, may produce some scatter in the left-hand panel of Figure\,\ref{fig:Fig2}, none of the observed ratios are as extreme as the $m_{\rm Na}/m_{\rm Ca}$ lower limit in NLTT\,5306.

At face value, this highly non-solar ratio implies that the outer layers of the brown dwarf are depleted in Ca, which acts as an empirical indicator that the atmosphere of NLTT\,5306B is differentiated (as expected) and distinct from a low-mass star. Speculating somewhat, and assuming the brown dwarf atmosphere is cloudy, Ca could be locked up in molecules as simple as CaH, or perovskite (CaTiO$_3$), which starts condensing below 1700\,K, forming clouds \citep{1996Fegley,2001Burrows,2002Lodders}.  In contrast, Na would not be locked up in any molecule and should remain atomic, consistent with spectroscopy and atmospheric modelling appropriate for L dwarf temperatures \citep{1999Kirkpatrick,1999Lodders,2005Cushing,2017Line}. At least superficially, these simple expectations of brown dwarf atmospheres are consistent with the data for NLTT\,5306.

\subsubsection{On the hypothesized inflation and magnetism}

Before further discussion, previous work and assertions made for NLTT\,5306 are critically examined.  In particular, it has been claimed that the brown dwarf is inflated, and that white dwarf magnetism plays a fundamental role in the binary properties (e.g.\ mass transfer).

At first glance, an inflated radius may have relevance to the detection of mass loss.  However, the single infrared spectrum on which the inference is based is not only model dependent (from where the gravity indicators originate), but also sensitive to the accuracy of the modelled and subtracted white dwarf photosphere at infrared wavelengths \citep{2020Casewell_b}.  Apart from the inferred shape of its infrared spectrum, there is no empirical support for an inflated radius in NLTT\,5306B. 

In contrast, WD\,1032+011 is a similar (white dwarf + brown dwarf) system, with a radius determined from eclipse measurements, but even in this case, the radius is within $2\upsigma-3\upsigma$ of several reasonable models \citep{2020Casewell_a}, where the binary can be younger than suggested by single-star kinematics\footnote{Any one star of any given age can have any space motion; it is only the dispersions exhibited by populations that yield deterministic ages based on non-kinematically selected benchmark populations (e.g.\ \citealt{2004Nordstrom,2014Bensby}).}.  But more importantly, there is no suggestion of mass transfer in the WD\,1032+011 system, despite indications of an enlarged brown dwarf radius.  There are a small number of eclipsing brown dwarf companions to white dwarfs, and only in the case of WD\,1032+011 is there an indication of inflation \citep{2014Littlefair,2017Parsons_a}.  Perhaps similarly, there is a well-documented scatter in radii for M dwarfs which extends down to 90\,M$_{\rm Jup}$ \citep{2018Parsons}.  As these radii are derived primarily via eclipses with both main-sequence and white dwarf primaries, the cause cannot be any influence from a white dwarf host, and L-type brown dwarfs may simply exhibit a similar pattern (see next Section).  

In the case of NLTT\,5306B, for a 55\,M$_{\rm Jup}$ brown dwarf of age greater than 1\,Gyr, evolutionary models predict the radius should be smaller than 0.096\,R$_\odot$, whereas the Roche lobe is 0.125\,R$_\odot$.  While neither the age nor mass are tightly constrained, for any reasonable range of parameters, the brown dwarf should be far from filling its Roche lobe \citep{2015Baraffe}.  Moreover, as previously mentioned, if NLTT\,5306B were abnormally enlarged and {\em transferring} mass gravitationally, the accreted $m_{\rm Na}/m_{\rm Ca}$ ratio should be mundane, and not the anomalous value observed. Lastly, a wind Roche lobe overflow mechanism exists which lies in between the standard Roche lobe overflow and Bondi-Hoyle accretion \citep{2007Mohamed,2009deValBorro}, but has so far only been applied to AGB stars with dusty and massive winds, and requires a significant wind acceleration zone.  AGB wind speeds range from 5 to 30\,km\,s$^{-1}$ \citep{2014Mayer,2017Goldman}, and are at least an order of magnitude slower than the wind speeds expected of M and L dwarfs, and that observed from the Sun.  Thus, it is currently unclear if wind Roche lobe overflow applies to mass loss in main-sequence stars, and further research would clarify if the flow diverges significantly from Bondi-Hoyle.

Separately but related, previous work on NLTT\,5306 has repeatedly suggested that the white dwarf might be weakly magnetic, with a field strength below the currently detectable threshold of Zeeman splitting in the optical \citep{2019Longstaff,2020Casewell_b,2022Buzard}. The motivation for a putative white dwarf magnetism appears to be twofold: to account for the mass accretion onto the white dwarf, and to sustain the claimed brown dwarf inflation.

First, magnetic funnelling of infalling material was inferred for NLTT\,5306 based on "the lack of any observational evidence for a substantial accretion disc" \citep{2019Longstaff}, but the same is true for non-magnetic white dwarfs accreting stellar wind from their M dwarf companions.  All eight white dwarf + M dwarf pairs in Table\,\ref{tab:tab1} have neither detected magnetic fields (via Zeeman splitting or cyclotron emission) nor any indications of accretion discs, yet are nevertheless profusely polluted from intrinsic mass loss from their companions.  The same can be true for NLTT\,5306.

Second, in the case of NLTT\,5306, there has been speculation regarding the potential role of magnetic interactions between the brown dwarf and the primary star, although no specific models or underlying physical mechanism have been proposed \citep{2020Casewell_b}. A follow-up study searched for a high metallicity and cloudy brown dwarf atmosphere that might result in a sizeable radius enhancement \citep{2011Burrows}, but was discounted in favour of inflation via white dwarf magnetism, albeit without referencing models or prototypes where such phenomena have been observed \citep{2022Buzard}.

It is not clear if there are any existing models or prototypical examples of magnetic white dwarfs that cause stellar or substellar companions to become inflated.  A strongly magnetic white dwarf ($B\sim100$\,MG) might heat an asteroid or possibly a dwarf planet-sized body via Ohmic dissipation \citep{2019Bromley}, but the energy required to inflate a brown dwarf is many orders of magnitude larger.  Brown dwarfs are degenerate and have the highest density of unevolved astrophysical objects (not stellar remnants), and to inflate the radius requires an energy source comparable to their internal energy.  To accelerate NLTT\,5306B to its current rotation rate (from zero) requires around 1 per cent of its gravitational potential, and, assuming there is actually any inflation, tidal synchronization may be better suited to the task, but this is speculation.

\subsubsection{On the origin of the substellar mass loss}

In the previous sections, it was discussed that the accreted material cannot be from a debris disk or a case of Roche lobe overflow based on the material composition. This conclusion is robust even against the pathological cases that the white dwarf atmosphere is observed either in the increasing or decreasing phase of accretion.  In the increasing phase where accretion has been ongoing for less than a sinking timescale of roughly $10^4$\,yr, the material being accreted would have $m_{\rm Na}/m_{\rm Ca} > 1000$ by mass.  If the system is observed in the decreasing phase so that the material originally had solar (or chondritic) abundance, then accretion should have ceased approximately 1.1\,Myr prior, or roughly 20 sinking timescales for Na and nearly 30 timescales for Ca. This implies unphysical masses of pollutants (e.g.\ $10^{20}$ times the mass of Na observed today).

It can also be reasoned that the interstellar medium cannot be the source of the observed Na pollution in NLTT\,5306. First, while highly refractory Ca may be locked up in dust grains more often than Na in the interstellar medium, the extreme ratio is incompatible, as Ca will sublimate into a gas at sufficiently close distances to the white dwarf and accrete. Second, the H$\upalpha$ emission line would have to be supported by a sufficiently high interstellar accretion rate, and it would be expected in other stars but has only been seen in a small number of white dwarfs with close companions where wind accretion is suspected or known \citep{2003Zuckerman, 2006Debes, 2011Tappert_a}. Furthermore, the density of the interstellar medium implied by the observed accretion rate onto the white dwarf, based on its tangential speed, suggests densities higher than observed in a molecular cloud.  In accordance with these reasons, the interstellar medium can be discounted as a viable source of Na pollution.

All data indicate the white dwarf is currently accreting in a steady state from a source with a high degree of chemical differentiation, as observed between Na and Ca. A plausible source for this material is the outermost layers of the brown dwarf, lost in an intrinsic wind, which acts as the source of pollution. The immediate cause of this differentiated wind is unknown, but by default is presumed to be the acceleration of Na ions along the magnetic field lines of the rapidly rotating brown dwarf.  To account for this inference, there must be a mechanism to preferentially ionise Na atoms. The atmospheric chemistry of brown dwarfs certainly plays a crucial role, with Ca predicted to be locked up in molecules or cloud particles, whereas Na will be in an atomic state.

In this picture, the alkali metals Li and K might be similarly expected, as they should remain atomic for a range of the warmest dwarf temperatures \citep{1999Lodders,2001Burrows}.  The co-added spectrum of NLTT\,5306 does not exhibit absorption from these elements, where the estimated upper limits [Li/H] $< -8.5$ and [K/H] $ < -7.0$ translate into abundance upper limits in the captured wind of $m_{\rm Li}/m_{\rm Na} < 0.03$ and $m_{\rm K}/m_{\rm Na} < 10$.  Unfortunately, both these limits are two orders of magnitude higher than the solar values \citep{2003Lodders}, and thus no further inferences can be drawn at present.

Grounded in the detection of Na in one case, one might expect analogous pollution in the other three white dwarfs with similar brown dwarf companions in close orbit.  However, as mentioned above and shown in Table\,\ref{tab:tab1} those observations are {\em insensitive to the Na abundance detected for NLTT\,5306}.  Future observations of physically similar systems, but with more favourable detection sensitivities, may be able to uncover analogous signs of chemically-peculiar pollution from differentiated winds in ultracool dwarfs.

A priori, isolated brown dwarfs are unlikely to have significant wind due to their low atmospheric ionisation levels, resulting in weak atmosphere-field coupling \citep{2002Mohanty}. However, in a post-common envelope binary system, the white dwarf may raise the ionisation fraction in the brown dwarf atmosphere via irradiation, potentially facilitating a stronger flow. To determine the contribution of ionisation to wind liberation, a recombination-limited case was considered using an appropriate DA model white dwarf spectrum (following the methodology from \citealt{2016Owen_a}). Assuming the recombination limit, the resulting maximum mass outflow rate was estimated to be $10^{-23}$\,\mdot, drastically lower than the empirically estimated mass-loss rate. Given the trivial ionising photon flux in a 7700\,K white dwarf, the recombination limit approximation is unlikely to hold. However, the correct energy-limited value would be even lower, thus making no qualitative difference. 

Other externally-driven mass loss mechanisms were considered for NLTT\,5306B but discounted.  Non-thermal mass loss and Jeans escape are negligible, and atmospheric boil-off \citep{2016Owen_b} would require the brown dwarf to significantly overfill its Roche lobe and is hence ruled out.  The mass loss from the brown dwarf caused by X-ray and extreme ultraviolet ionising radiation emitted by the white dwarf \citep{2007Erkaev} is also calculated to be negligible for all four systems studied here. In summary, the radiation fields of these cool white dwarfs should have no significant effect on the coronal activity of the brown dwarf.

\section{Conclusions}

The goal of this work is to obtain sensitive limits to mass loss via substellar winds using white dwarf hosts.  Using archival data for four known white dwarfs with closely-orbiting brown dwarf companions, upper limits for intrinsic mass loss are found to be three orders of magnitude lower than the solar value, concluded from the absence of Ca\,{\sc ii}\,K absorption. On the one hand, this methodology demonstrates similar sensitivity to the mass-loss estimates obtained from Ly$\upalpha$ observations of M dwarfs, but only requires optical spectra and is not limited to bright stars within 7\,pc.  On the other hand, only post-common envelope binaries with a (polluted) white dwarf can be studied this way.  The brown dwarf mass-loss limits are then compared to M dwarfs for which abundances of at least two elements are known. It is found that L dwarf wind upper limits are approximately an order of magnitude below the $\dot M$ values calculated for the M dwarfs. 

The sole exception is the detection of Na in NLTT\,5306, where a white dwarf accretion rate and corresponding L dwarf mass-loss rate are estimated, but uncertain owing to the highly unusual composition.  The accreted material is shown to have $m_{\rm Na}/m_{\rm Ca}>900$ based on the Na detection and Ca upper limit (cf.\ [$m_{\rm Na}/m_{\rm Ca}]_\odot\approx1$). This extreme ratio rules out planetary debris of any known origin, as well as the interstellar medium.  Notably, the composition appears consistent with atomic abundance predictions for a warm brown dwarf upper atmosphere.  Derived from such models, the alkali elements Li and K might also be incorporated into a wind \citep{2017Line,2021Gharib}, but the observations are currently insensitive to these elements at solar ratios.  This differentiated material cannot have reached the white dwarf via Roche lobe overflow, and instead an intrinsic wind must be the source of pollution, making it the first detection of mass loss in a substellar object.  An important caveat is that the 1.7\,h spin period of NLTT\,5306B may suggest that its substellar wind is not representative of most L dwarfs.

This study demonstrates that the complex atmospheres of brown dwarfs, which may reflect multiple formation pathways and chemical diversity \citep{2013Apai, 2014Helling,2016Madhusudhan}, can be probed to some degree using polluted white dwarf hosts.  Importantly, it is demonstrated that the white dwarf hosts cannot radiatively induce significant mass loss in their companions.  However, some intrinsic binary features are not necessarily representative of all isolated ultracool dwarfs, such as rapid rotation, and may influence mass loss.  At present, these effects are difficult to quantify with the existing data.  Further work on the captured wind flow would better constrain low-mass stellar and substellar mass-loss rates, and a sensitive search for Na, Li, and K in cooler white dwarfs with ultracool dwarf companions could prove fruitful (e.g.\ WISEA J061543.91$-$124726.8; \citealt{2016Fajardo}).

\section*{Acknowledgements}
The authors acknowledge L.~Fossati, S.~Mohanty, and J.~E.~Owen for calculations related to the irradiation of brown dwarf atmospheres, S.~G.~Parsons for input on possible physical models for the inflation of low-mass companions to white dwarfs, M.~S.~Marley and A.~Burrows for discussions on ultracool dwarf abundances, D.~Koester for a white dwarf atmospheric model that includes the extreme ultraviolet, D.~Hendriks for helpful feedback on the manuscript, and the referee for useful comments that improved the final manuscript.  This work is based on archival data from the European Southern Observatory Very Large Telescope and Science Archive Facility as well as data collected by the {\em TESS} mission publicly available from the Mikulski Archive for Space Telescopes. The analysis utilised the Montreal White Dwarf Database\footnote{\url{https://www.montrealwhitedwarfdatabase.org/home.html}}, white dwarf atmospheric models made available D. Koester\footnote{\url{http://www1.astrophysik.uni-kiel.de/~koester/astrophysics/astrophysics.html}}, and brown dwarf evolutionary models provided by I. Baraffe\footnote{\url{http://perso.ens-lyon.fr/isabelle.baraffe/BHAC15dir/BHAC15_tracks+structure}}. NW was supported by a UK STFC studentship hosted by the UCL Centre for Doctoral Training in Data Intensive Science.  JF acknowledges STFC grant ST/R000476/1.

\section*{Data Availability}

The data underlying this article is publicly available.

%\begin{thebibliography}{}

\bibliographystyle{mnras}
\bibliography{references}

\begin{thebibliography}{}
\makeatletter
\relax
\def\mn@urlcharsother{\let\do\@makeother \do\$\do\&\do\#\do\^\do\_\do\%\do\~}
\def\mn@doi{\begingroup\mn@urlcharsother \@ifnextchar [ {\mn@doi@}
  {\mn@doi@[]}}
\def\mn@doi@[#1]#2{\def\@tempa{#1}\ifx\@tempa\@empty \href
  {http://dx.doi.org/#2} {doi:#2}\else \href {http://dx.doi.org/#2} {#1}\fi
  \endgroup}
\def\mn@eprint#1#2{\mn@eprint@#1:#2::\@nil}
\def\mn@eprint@arXiv#1{\href {http://arxiv.org/abs/#1} {{\tt arXiv:#1}}}
\def\mn@eprint@dblp#1{\href {http://dblp.uni-trier.de/rec/bibtex/#1.xml}
  {dblp:#1}}
\def\mn@eprint@#1:#2:#3:#4\@nil{\def\@tempa {#1}\def\@tempb {#2}\def\@tempc
  {#3}\ifx \@tempc \@empty \let \@tempc \@tempb \let \@tempb \@tempa \fi \ifx
  \@tempb \@empty \def\@tempb {arXiv}\fi \@ifundefined
  {mn@eprint@\@tempb}{\@tempb:\@tempc}{\expandafter \expandafter \csname
  mn@eprint@\@tempb\endcsname \expandafter{\@tempc}}}

\bibitem[\protect\citeauthoryear{{All{\`e}gre}, {Poirier}, {Humler}  \&
  {Hofmann}}{{All{\`e}gre} et~al.}{1995}]{1995Allegre}
{All{\`e}gre} C.~J.,  {Poirier} J.-P.,  {Humler} E.,   {Hofmann} A.~W.,  1995,
  \mn@doi [Earth and Planetary Science Letters] {10.1016/0012-821X(95)00123-T},
  \href {https://ui.adsabs.harvard.edu/abs/1995E&PSL.134..515A} {134, 515}

\bibitem[\protect\citeauthoryear{{Amaro} et~al.,}{{Amaro}
  et~al.}{2023}]{2023Amaro}
{Amaro} R.~C.,  et~al., 2023, arXiv e-prints, \href
  {https://ui.adsabs.harvard.edu/abs/2023arXiv230307420A} {p. arXiv:2303.07420}

\bibitem[\protect\citeauthoryear{{Apai}, {Radigan}, {Buenzli}, {Burrows},
  {Reid}  \& {Jayawardhana}}{{Apai} et~al.}{2013}]{2013Apai}
{Apai} D.,  {Radigan} J.,  {Buenzli} E.,  {Burrows} A.,  {Reid} I.~N.,
  {Jayawardhana} R.,  2013, \mn@doi [\apj] {10.1088/0004-637X/768/2/121}, \href
  {https://ui.adsabs.harvard.edu/abs/2013ApJ...768..121A} {768, 121}

\bibitem[\protect\citeauthoryear{{Ashley}, {Farihi}, {Marsh}, {Wilson}  \&
  {G{\"a}nsicke}}{{Ashley} et~al.}{2019}]{2019Ashley}
{Ashley} R.~P.,  {Farihi} J.,  {Marsh} T.~R.,  {Wilson} D.~J.,   {G{\"a}nsicke}
  B.~T.,  2019, \mn@doi [\mnras] {10.1093/mnras/stz298}, \href
  {https://ui.adsabs.harvard.edu/abs/2019MNRAS.484.5362A} {484, 5362}

\bibitem[\protect\citeauthoryear{{Bagnulo} \& {Landstreet}}{{Bagnulo} \&
  {Landstreet}}{2021}]{2021Bagnulo}
{Bagnulo} S.,  {Landstreet} J.~D.,  2021, \mn@doi [\mnras]
  {10.1093/mnras/stab2046}, \href
  {https://ui.adsabs.harvard.edu/abs/2021MNRAS.507.5902B} {507, 5902}

\bibitem[\protect\citeauthoryear{{Ballester}, {Modigliani}, {Boitquin},
  {Cristiani}, {Hanuschik}, {Kaufer}  \& {Wolf}}{{Ballester}
  et~al.}{2000}]{2000Ballester}
{Ballester} P.,  {Modigliani} A.,  {Boitquin} O.,  {Cristiani} S.,  {Hanuschik}
  R.,  {Kaufer} A.,   {Wolf} S.,  2000, The Messenger, \href
  {https://ui.adsabs.harvard.edu/abs/2000Msngr.101...31B} {101, 31}

\bibitem[\protect\citeauthoryear{{Baraffe}, {Homeier}, {Allard}  \&
  {Chabrier}}{{Baraffe} et~al.}{2015}]{2015Baraffe}
{Baraffe} I.,  {Homeier} D.,  {Allard} F.,   {Chabrier} G.,  2015, \mn@doi
  [\aap] {10.1051/0004-6361/201425481}, \href
  {https://ui.adsabs.harvard.edu/abs/2015A&A...577A..42B} {577, A42}

\bibitem[\protect\citeauthoryear{{Barstow}, {Barstow}, {Casewell}, {Holberg}
  \& {Hubeny}}{{Barstow} et~al.}{2014}]{2014Barstow}
{Barstow} M.~A.,  {Barstow} J.~K.,  {Casewell} S.~L.,  {Holberg} J.~B.,
  {Hubeny} I.,  2014, \mn@doi [\mnras] {10.1093/mnras/stu216}, \href
  {https://ui.adsabs.harvard.edu/abs/2014MNRAS.440.1607B} {440, 1607}

\bibitem[\protect\citeauthoryear{{Bensby}, {Feltzing}  \& {Oey}}{{Bensby}
  et~al.}{2014}]{2014Bensby}
{Bensby} T.,  {Feltzing} S.,   {Oey} M.~S.,  2014, \mn@doi [\aap]
  {10.1051/0004-6361/201322631}, \href
  {https://ui.adsabs.harvard.edu/abs/2014A&A...562A..71B} {562, A71}

\bibitem[\protect\citeauthoryear{{Berger} et~al.,}{{Berger}
  et~al.}{2009}]{2009Berger}
{Berger} E.,  et~al., 2009, \mn@doi [\apj] {10.1088/0004-637X/695/1/310}, \href
  {https://ui.adsabs.harvard.edu/abs/2009ApJ...695..310B} {695, 310}

\bibitem[\protect\citeauthoryear{{Beuermann} et~al.,}{{Beuermann}
  et~al.}{2013}]{2013Beuermann}
{Beuermann} K.,  et~al., 2013, \mn@doi [\aap] {10.1051/0004-6361/201322241},
  \href {https://ui.adsabs.harvard.edu/abs/2013A&A...558A..96B} {558, A96}

\bibitem[\protect\citeauthoryear{{Blake}, {Charbonneau}  \& {White}}{{Blake}
  et~al.}{2010}]{2010Blake}
{Blake} C.~H.,  {Charbonneau} D.,   {White} R.~J.,  2010, \mn@doi [\apj]
  {10.1088/0004-637X/723/1/684}, \href
  {https://ui.adsabs.harvard.edu/abs/2010ApJ...723..684B} {723, 684}

\bibitem[\protect\citeauthoryear{{Bondi}}{{Bondi}}{1952}]{1952Bondi}
{Bondi} H.,  1952, \mn@doi [\mnras] {10.1093/mnras/112.2.195}, \href
  {https://ui.adsabs.harvard.edu/abs/1952MNRAS.112..195B} {112, 195}

\bibitem[\protect\citeauthoryear{{Bromley} \& {Kenyon}}{{Bromley} \&
  {Kenyon}}{2019}]{2019Bromley}
{Bromley} B.~C.,  {Kenyon} S.~J.,  2019, \mn@doi [\apj]
  {10.3847/1538-4357/ab12e9}, \href
  {https://ui.adsabs.harvard.edu/abs/2019ApJ...876...17B} {876, 17}

\bibitem[\protect\citeauthoryear{{Burleigh}, {Hogan}, {Dobbie}, {Napiwotzki}
  \& {Maxted}}{{Burleigh} et~al.}{2006}]{2006Burleigh}
{Burleigh} M.~R.,  {Hogan} E.,  {Dobbie} P.~D.,  {Napiwotzki} R.,   {Maxted}
  P.~F.~L.,  2006, \mn@doi [\mnras] {10.1111/j.1745-3933.2006.00242.x}, \href
  {https://ui.adsabs.harvard.edu/abs/2006MNRAS.373L..55B} {373, L55}

\bibitem[\protect\citeauthoryear{{Burrows}, {Hubbard}, {Lunine}  \&
  {Liebert}}{{Burrows} et~al.}{2001}]{2001Burrows}
{Burrows} A.,  {Hubbard} W.~B.,  {Lunine} J.~I.,   {Liebert} J.,  2001, \mn@doi
  [Reviews of Modern Physics] {10.1103/RevModPhys.73.719}, \href
  {https://ui.adsabs.harvard.edu/abs/2001RvMP...73..719B} {73, 719}

\bibitem[\protect\citeauthoryear{{Burrows}, {Heng}  \& {Nampaisarn}}{{Burrows}
  et~al.}{2011}]{2011Burrows}
{Burrows} A.,  {Heng} K.,   {Nampaisarn} T.,  2011, \mn@doi [\apj]
  {10.1088/0004-637X/736/1/47}, \href
  {https://ui.adsabs.harvard.edu/abs/2011ApJ...736...47B} {736, 47}

\bibitem[\protect\citeauthoryear{{Buzard}, {Casewell}, {Lothringer}  \&
  {Blake}}{{Buzard} et~al.}{2022}]{2022Buzard}
{Buzard} C.,  {Casewell} S.,  {Lothringer} J.,   {Blake} G.,  2022, arXiv
  e-prints, \href {https://ui.adsabs.harvard.edu/abs/2022arXiv220405330B} {p.
  arXiv:2204.05330}

\bibitem[\protect\citeauthoryear{{Casewell}, {Littlefair}, {Parsons}, {Marsh},
  {Fortney}  \& {Marley}}{{Casewell} et~al.}{2018}]{2018Casewell_a}
{Casewell} S.~L.,  {Littlefair} S.~P.,  {Parsons} S.~G.,  {Marsh} T.~R.,
  {Fortney} J.~J.,   {Marley} M.~S.,  2018, \mn@doi [\mnras]
  {10.1093/mnras/sty2599}, \href
  {https://ui.adsabs.harvard.edu/abs/2018MNRAS.481.5216C} {481, 5216}

\bibitem[\protect\citeauthoryear{{Casewell} et~al.,}{{Casewell}
  et~al.}{2020a}]{2020Casewell_a}
{Casewell} S.~L.,  et~al., 2020a, \mn@doi [\mnras] {10.1093/mnras/staa1608},
  \href {https://ui.adsabs.harvard.edu/abs/2020MNRAS.497.3571C} {497, 3571}

\bibitem[\protect\citeauthoryear{{Casewell}, {Debes}, {Braker}, {Cushing},
  {Mace}, {Marley}  \& {Kirkpatrick}}{{Casewell}
  et~al.}{2020b}]{2020Casewell_b}
{Casewell} S.~L.,  {Debes} J.,  {Braker} I.~P.,  {Cushing} M.~C.,  {Mace} G.,
  {Marley} M.~S.,   {Kirkpatrick} J.~D.,  2020b, \mn@doi [\mnras]
  {10.1093/mnras/staa3184}, \href
  {https://ui.adsabs.harvard.edu/abs/2020MNRAS.499.5318C} {499, 5318}

\bibitem[\protect\citeauthoryear{{Chayer}, {Fontaine}  \& {Wesemael}}{{Chayer}
  et~al.}{1995}]{1995Chayer}
{Chayer} P.,  {Fontaine} G.,   {Wesemael} F.,  1995, \mn@doi [\apjs]
  {10.1086/192184}, \href
  {https://ui.adsabs.harvard.edu/abs/1995ApJS...99..189C} {99, 189}

\bibitem[\protect\citeauthoryear{{Christensen}, {Holzwarth}  \&
  {Reiners}}{{Christensen} et~al.}{2009}]{2009Christensen}
{Christensen} U.~R.,  {Holzwarth} V.,   {Reiners} A.,  2009, \mn@doi [\nat]
  {10.1038/nature07626}, \href
  {https://ui.adsabs.harvard.edu/abs/2009Natur.457..167C} {457, 167}

\bibitem[\protect\citeauthoryear{{Cohen}}{{Cohen}}{2011}]{2011Cohen}
{Cohen} O.,  2011, \mn@doi [\mnras] {10.1111/j.1365-2966.2011.19428.x}, \href
  {https://ui.adsabs.harvard.edu/abs/2011MNRAS.417.2592C} {417, 2592}

\bibitem[\protect\citeauthoryear{{Cook}, {Williams}  \& {Berger}}{{Cook}
  et~al.}{2014}]{2014Cook}
{Cook} B.~A.,  {Williams} P.~K.~G.,   {Berger} E.,  2014, \mn@doi [\apj]
  {10.1088/0004-637X/785/1/10}, \href
  {https://ui.adsabs.harvard.edu/abs/2014ApJ...785...10C} {785, 10}

\bibitem[\protect\citeauthoryear{{Coutu}, {Dufour}, {Bergeron}, {Blouin},
  {Loranger}, {Allard}  \& {Dunlap}}{{Coutu} et~al.}{2019}]{2019Coutu}
{Coutu} S.,  {Dufour} P.,  {Bergeron} P.,  {Blouin} S.,  {Loranger} E.,
  {Allard} N.~F.,   {Dunlap} B.~H.,  2019, \mn@doi [\apj]
  {10.3847/1538-4357/ab46b9}, \href
  {https://ui.adsabs.harvard.edu/abs/2019ApJ...885...74C} {885, 74}

\bibitem[\protect\citeauthoryear{{Cranmer} \& {Saar}}{{Cranmer} \&
  {Saar}}{2011}]{2011Cranmer}
{Cranmer} S.~R.,  {Saar} S.~H.,  2011, \mn@doi [\apj]
  {10.1088/0004-637X/741/1/54}, \href
  {https://ui.adsabs.harvard.edu/abs/2011ApJ...741...54C} {741, 54}

\bibitem[\protect\citeauthoryear{{Cushing}, {Rayner}  \& {Vacca}}{{Cushing}
  et~al.}{2005}]{2005Cushing}
{Cushing} M.~C.,  {Rayner} J.~T.,   {Vacca} W.~D.,  2005, \mn@doi [\apj]
  {10.1086/428040}, \href
  {https://ui.adsabs.harvard.edu/abs/2005ApJ...623.1115C} {623, 1115}

\bibitem[\protect\citeauthoryear{{Debes}}{{Debes}}{2006}]{2006Debes}
{Debes} J.~H.,  2006, \mn@doi [\apj] {10.1086/508132}, \href
  {https://ui.adsabs.harvard.edu/abs/2006ApJ...652..636D} {652, 636}

\bibitem[\protect\citeauthoryear{{Dekker}, {D'Odorico}, {Kaufer}, {Delabre}  \&
  {Kotzlowski}}{{Dekker} et~al.}{2000}]{2000Dekker}
{Dekker} H.,  {D'Odorico} S.,  {Kaufer} A.,  {Delabre} B.,   {Kotzlowski} H.,
  2000, in {Iye} M.,  {Moorwood} A.~F.,  eds,  Society of Photo-Optical
  Instrumentation Engineers (SPIE) Conference Series Vol. 4008, Optical and IR
  Telescope Instrumentation and Detectors. pp 534--545,
  \mn@doi{10.1117/12.395512}

\bibitem[\protect\citeauthoryear{{Doschek}, {Warren}  \& {Feldman}}{{Doschek}
  et~al.}{2015}]{2015Doschek}
{Doschek} G.~A.,  {Warren} H.~P.,   {Feldman} U.,  2015, \mn@doi [\apjl]
  {10.1088/2041-8205/808/1/L7}, \href
  {https://ui.adsabs.harvard.edu/abs/2015ApJ...808L...7D} {808, L7}

\bibitem[\protect\citeauthoryear{{Dufour}, {Kilic}, {Fontaine}, {Bergeron},
  {Melis}  \& {Bochanski}}{{Dufour} et~al.}{2012}]{2012Dufour}
{Dufour} P.,  {Kilic} M.,  {Fontaine} G.,  {Bergeron} P.,  {Melis} C.,
  {Bochanski} J.,  2012, \mn@doi [\apj] {10.1088/0004-637X/749/1/6}, \href
  {https://ui.adsabs.harvard.edu/abs/2012ApJ...749....6D} {749, 6}

\bibitem[\protect\citeauthoryear{{Dupuis}, {Fontaine}, {Pelletier}  \&
  {Wesemael}}{{Dupuis} et~al.}{1992}]{1992Dupuis}
{Dupuis} J.,  {Fontaine} G.,  {Pelletier} C.,   {Wesemael} F.,  1992, \mn@doi
  [\apjs] {10.1086/191728}, \href
  {https://ui.adsabs.harvard.edu/abs/1992ApJS...82..505D} {82, 505}

\bibitem[\protect\citeauthoryear{{Dupuis}, {Fontaine}, {Pelletier}  \&
  {Wesemael}}{{Dupuis} et~al.}{1993a}]{1993aDupuis}
{Dupuis} J.,  {Fontaine} G.,  {Pelletier} C.,   {Wesemael} F.,  1993a, \mn@doi
  [\apjs] {10.1086/191746}, \href
  {https://ui.adsabs.harvard.edu/abs/1993ApJS...84...73D} {84, 73}

\bibitem[\protect\citeauthoryear{{Dupuis}, {Fontaine}  \& {Wesemael}}{{Dupuis}
  et~al.}{1993b}]{1993bDupuis}
{Dupuis} J.,  {Fontaine} G.,   {Wesemael} F.,  1993b, \mn@doi [\apjs]
  {10.1086/191808}, \href
  {https://ui.adsabs.harvard.edu/abs/1993ApJS...87..345D} {87, 345}

\bibitem[\protect\citeauthoryear{{Eddington}}{{Eddington}}{1926}]{1926Eddington}
{Eddington} A.~S.,  1926, {The Internal Constitution of the Stars}.
Cambridge Science Classics, Cambridge University Press,
  \mn@doi{10.1017/CBO9780511600005}

\bibitem[\protect\citeauthoryear{{Erkaev}, {Kulikov}, {Lammer}, {Selsis},
  {Langmayr}, {Jaritz}  \& {Biernat}}{{Erkaev} et~al.}{2007}]{2007Erkaev}
{Erkaev} N.~V.,  {Kulikov} Y.~N.,  {Lammer} H.,  {Selsis} F.,  {Langmayr} D.,
  {Jaritz} G.~F.,   {Biernat} H.~K.,  2007, \mn@doi [\aap]
  {10.1051/0004-6361:20066929}, \href
  {https://ui.adsabs.harvard.edu/abs/2007A&A...472..329E} {472, 329}

\bibitem[\protect\citeauthoryear{{Fajardo-Acosta} et~al.,}{{Fajardo-Acosta}
  et~al.}{2016}]{2016Fajardo}
{Fajardo-Acosta} S.~B.,  et~al., 2016, \mn@doi [\apj]
  {10.3847/0004-637X/832/1/62}, \href
  {https://ui.adsabs.harvard.edu/abs/2016ApJ...832...62F} {832, 62}

\bibitem[\protect\citeauthoryear{{Farihi}}{{Farihi}}{2016}]{2016Farihi}
{Farihi} J.,  2016, \mn@doi [\nar] {10.1016/j.newar.2016.03.001}, \href
  {https://ui.adsabs.harvard.edu/abs/2016NewAR..71....9F} {71, 9}

\bibitem[\protect\citeauthoryear{{Farihi} \& {Christopher}}{{Farihi} \&
  {Christopher}}{2004}]{2004Farihi}
{Farihi} J.,  {Christopher} M.,  2004, \mn@doi [\aj] {10.1086/423919}, \href
  {https://ui.adsabs.harvard.edu/abs/2004AJ....128.1868F} {128, 1868}

\bibitem[\protect\citeauthoryear{{Farihi}, {Becklin}  \& {Zuckerman}}{{Farihi}
  et~al.}{2005}]{2005Farihi_a}
{Farihi} J.,  {Becklin} E.~E.,   {Zuckerman} B.,  2005, \mn@doi [\apjs]
  {10.1086/444362}, \href
  {https://ui.adsabs.harvard.edu/abs/2005ApJS..161..394F} {161, 394}

\bibitem[\protect\citeauthoryear{{Farihi}, {Burleigh}  \& {Hoard}}{{Farihi}
  et~al.}{2008}]{2008Farihi}
{Farihi} J.,  {Burleigh} M.~R.,   {Hoard} D.~W.,  2008, \mn@doi [\apj]
  {10.1086/524933}, \href
  {https://ui.adsabs.harvard.edu/abs/2008ApJ...674..421F} {674, 421}

\bibitem[\protect\citeauthoryear{{Farihi}, {Jura}  \& {Zuckerman}}{{Farihi}
  et~al.}{2009}]{2009Farihi}
{Farihi} J.,  {Jura} M.,   {Zuckerman} B.,  2009, \mn@doi [\apj]
  {10.1088/0004-637X/694/2/805}, \href
  {https://ui.adsabs.harvard.edu/abs/2009ApJ...694..805F} {694, 805}

\bibitem[\protect\citeauthoryear{{Farihi}, {Hoard}  \& {Wachter}}{{Farihi}
  et~al.}{2010a}]{2010Farihi_b}
{Farihi} J.,  {Hoard} D.~W.,   {Wachter} S.,  2010a, \mn@doi [\apjs]
  {10.1088/0067-0049/190/2/275}, \href
  {https://ui.adsabs.harvard.edu/abs/2010ApJS..190..275F} {190, 275}

\bibitem[\protect\citeauthoryear{{Farihi}, {Barstow}, {Redfield}, {Dufour}  \&
  {Hambly}}{{Farihi} et~al.}{2010b}]{2010Farihi_a}
{Farihi} J.,  {Barstow} M.~A.,  {Redfield} S.,  {Dufour} P.,   {Hambly} N.~C.,
  2010b, \mn@doi [\mnras] {10.1111/j.1365-2966.2010.16426.x}, \href
  {https://ui.adsabs.harvard.edu/abs/2010MNRAS.404.2123F} {404, 2123}

\bibitem[\protect\citeauthoryear{{Fegley} \& {Lodders}}{{Fegley} \&
  {Lodders}}{1996}]{1996Fegley}
{Fegley} Bruce J.,  {Lodders} K.,  1996, \mn@doi [\apjl] {10.1086/310356},
  \href {https://ui.adsabs.harvard.edu/abs/1996ApJ...472L..37F} {472, L37}

\bibitem[\protect\citeauthoryear{{Fontaine} \& {Michaud}}{{Fontaine} \&
  {Michaud}}{1979}]{1979Fontaine}
{Fontaine} G.,  {Michaud} G.,  1979, \mn@doi [\apj] {10.1086/157247}, \href
  {https://ui.adsabs.harvard.edu/abs/1979ApJ...231..826F} {231, 826}

\bibitem[\protect\citeauthoryear{{Freudling}, {Romaniello}, {Bramich},
  {Ballester}, {Forchi}, {Garc{\'\i}a-Dabl{\'o}}, {Moehler}  \&
  {Neeser}}{{Freudling} et~al.}{2013}]{2013Freudling}
{Freudling} W.,  {Romaniello} M.,  {Bramich} D.~M.,  {Ballester} P.,  {Forchi}
  V.,  {Garc{\'\i}a-Dabl{\'o}} C.~E.,  {Moehler} S.,   {Neeser} M.~J.,  2013,
  \mn@doi [\aap] {10.1051/0004-6361/201322494}, \href
  {https://ui.adsabs.harvard.edu/abs/2013A&A...559A..96F} {559, A96}

\bibitem[\protect\citeauthoryear{{Gharib-Nezhad}, {Marley}, {Batalha},
  {Visscher}, {Freedman}  \& {Lupu}}{{Gharib-Nezhad} et~al.}{2021}]{2021Gharib}
{Gharib-Nezhad} E.,  {Marley} M.~S.,  {Batalha} N.~E.,  {Visscher} C.,
  {Freedman} R.~S.,   {Lupu} R.~E.,  2021, \mn@doi [\apj]
  {10.3847/1538-4357/ac0a7d}, \href
  {https://ui.adsabs.harvard.edu/abs/2021ApJ...919...21G} {919, 21}

\bibitem[\protect\citeauthoryear{{Gianninas}, {Bergeron}  \&
  {Ruiz}}{{Gianninas} et~al.}{2011}]{2011Gianninas}
{Gianninas} A.,  {Bergeron} P.,   {Ruiz} M.~T.,  2011, \mn@doi [\apj]
  {10.1088/0004-637X/743/2/138}, \href
  {https://ui.adsabs.harvard.edu/abs/2011ApJ...743..138G} {743, 138}

\bibitem[\protect\citeauthoryear{{Gizis}, {Monet}, {Reid}, {Kirkpatrick},
  {Liebert}  \& {Williams}}{{Gizis} et~al.}{2000}]{2000Gizis}
{Gizis} J.~E.,  {Monet} D.~G.,  {Reid} I.~N.,  {Kirkpatrick} J.~D.,  {Liebert}
  J.,   {Williams} R.~J.,  2000, \mn@doi [\aj] {10.1086/301456}, \href
  {https://ui.adsabs.harvard.edu/abs/2000AJ....120.1085G} {120, 1085}

\bibitem[\protect\citeauthoryear{{Goldman} et~al.,}{{Goldman}
  et~al.}{2017}]{2017Goldman}
{Goldman} S.~R.,  et~al., 2017, \mn@doi [\mnras] {10.1093/mnras/stw2708}, \href
  {https://ui.adsabs.harvard.edu/abs/2017MNRAS.465..403G} {465, 403}

\bibitem[\protect\citeauthoryear{{Hawley}, {Reid}  \& {Gizis}}{{Hawley}
  et~al.}{2000}]{2000Hawley}
{Hawley} S.,  {Reid} I.~N.,   {Gizis} J.,  2000, in {Griffith} C.~A.,  {Marley}
  M.~S.,  eds,  Astronomical Society of the Pacific Conference Series Vol. 212,
  From Giant Planets to Cool Stars. p.~252

\bibitem[\protect\citeauthoryear{{Helling} \& {Casewell}}{{Helling} \&
  {Casewell}}{2014}]{2014Helling}
{Helling} C.,  {Casewell} S.,  2014, \mn@doi [\aapr]
  {10.1007/s00159-014-0080-0}, \href
  {https://ui.adsabs.harvard.edu/abs/2014A&ARv..22...80H} {22, 80}

\bibitem[\protect\citeauthoryear{{H{\"o}fner} \& {Olofsson}}{{H{\"o}fner} \&
  {Olofsson}}{2018}]{2018Hofner}
{H{\"o}fner} S.,  {Olofsson} H.,  2018, \mn@doi [\aapr]
  {10.1007/s00159-017-0106-5}, \href
  {https://ui.adsabs.harvard.edu/abs/2018A&ARv..26....1H} {26, 1}

\bibitem[\protect\citeauthoryear{{Hollands}, {Koester}, {Alekseev}, {Herbert}
  \& {G{\"a}nsicke}}{{Hollands} et~al.}{2017}]{2017Hollands}
{Hollands} M.~A.,  {Koester} D.,  {Alekseev} V.,  {Herbert} E.~L.,
  {G{\"a}nsicke} B.~T.,  2017, \mn@doi [\mnras] {10.1093/mnras/stx250}, \href
  {https://ui.adsabs.harvard.edu/abs/2017MNRAS.467.4970H} {467, 4970}

\bibitem[\protect\citeauthoryear{{Hollands}, {Tremblay}, {G{\"a}nsicke},
  {Koester}  \& {Gentile-Fusillo}}{{Hollands} et~al.}{2021}]{2021Hollands}
{Hollands} M.~A.,  {Tremblay} P.-E.,  {G{\"a}nsicke} B.~T.,  {Koester} D.,
  {Gentile-Fusillo} N.~P.,  2021, \mn@doi [Nature Astronomy]
  {10.1038/s41550-020-01296-7}, \href
  {https://ui.adsabs.harvard.edu/abs/2021NatAs...5..451H} {5, 451}

\bibitem[\protect\citeauthoryear{{Hollands}, {Tremblay}, {G{\"a}nsicke}  \&
  {Koester}}{{Hollands} et~al.}{2022}]{2022Hollands}
{Hollands} M.~A.,  {Tremblay} P.~E.,  {G{\"a}nsicke} B.~T.,   {Koester} D.,
  2022, \mn@doi [\mnras] {10.1093/mnras/stab3696}, \href
  {https://ui.adsabs.harvard.edu/abs/2022MNRAS.511...71H} {511, 71}

\bibitem[\protect\citeauthoryear{{Joergens}, {Fern{\'a}ndez}, {Carpenter}  \&
  {Neuh{\"a}user}}{{Joergens} et~al.}{2003}]{2003Joergens}
{Joergens} V.,  {Fern{\'a}ndez} M.,  {Carpenter} J.~M.,   {Neuh{\"a}user} R.,
  2003, \mn@doi [\apj] {10.1086/377041}, \href
  {https://ui.adsabs.harvard.edu/abs/2003ApJ...594..971J} {594, 971}

\bibitem[\protect\citeauthoryear{{Johnstone}}{{Johnstone}}{2017}]{2017Johnstone}
{Johnstone} C.~P.,  2017, \mn@doi [\aap] {10.1051/0004-6361/201629609}, \href
  {https://ui.adsabs.harvard.edu/abs/2017A&A...598A..24J} {598, A24}

\bibitem[\protect\citeauthoryear{{Kafka}, {Tappert}  \& {Honeycutt}}{{Kafka}
  et~al.}{2010}]{2010Kafka}
{Kafka} S.,  {Tappert} C.,   {Honeycutt} R.~K.,  2010, \mn@doi [\mnras]
  {10.1111/j.1365-2966.2009.16063.x}, \href
  {https://ui.adsabs.harvard.edu/abs/2010MNRAS.403..755K} {403, 755}

\bibitem[\protect\citeauthoryear{{Kao}, {Hallinan}, {Pineda}, {Stevenson}  \&
  {Burgasser}}{{Kao} et~al.}{2018}]{2018Kao}
{Kao} M.~M.,  {Hallinan} G.,  {Pineda} J.~S.,  {Stevenson} D.,   {Burgasser}
  A.,  2018, \mn@doi [\apjs] {10.3847/1538-4365/aac2d5}, \href
  {https://ui.adsabs.harvard.edu/abs/2018ApJS..237...25K} {237, 25}

\bibitem[\protect\citeauthoryear{{Kawka}, {Vennes}, {Dupuis}, {Chayer}  \&
  {Lanz}}{{Kawka} et~al.}{2008}]{2008Kawka}
{Kawka} A.,  {Vennes} S.,  {Dupuis} J.,  {Chayer} P.,   {Lanz} T.,  2008,
  \mn@doi [\apj] {10.1086/526411}, \href
  {https://ui.adsabs.harvard.edu/abs/2008ApJ...675.1518K} {675, 1518}

\bibitem[\protect\citeauthoryear{{Khodachenko} et~al.,}{{Khodachenko}
  et~al.}{2007}]{2007Khodachenko}
{Khodachenko} M.~L.,  et~al., 2007, \mn@doi [Astrobiology]
  {10.1089/ast.2006.0127}, \href
  {https://ui.adsabs.harvard.edu/abs/2007AsBio...7..167K} {7, 167}

\bibitem[\protect\citeauthoryear{{Kirkpatrick} et~al.,}{{Kirkpatrick}
  et~al.}{1999}]{1999Kirkpatrick}
{Kirkpatrick} J.~D.,  et~al., 1999, \mn@doi [\apj] {10.1086/307414}, \href
  {https://ui.adsabs.harvard.edu/abs/1999ApJ...519..802K} {519, 802}

\bibitem[\protect\citeauthoryear{{Koester}}{{Koester}}{2009}]{2009Koester}
{Koester} D.,  2009, \mn@doi [\aap] {10.1051/0004-6361/200811468}, \href
  {https://ui.adsabs.harvard.edu/abs/2009A&A...498..517K} {498, 517}

\bibitem[\protect\citeauthoryear{{Koester} \& {Wilken}}{{Koester} \&
  {Wilken}}{2006}]{2006Koester}
{Koester} D.,  {Wilken} D.,  2006, \mn@doi [\aap] {10.1051/0004-6361:20064843},
  \href {https://ui.adsabs.harvard.edu/abs/2006A&A...453.1051K} {453, 1051}

\bibitem[\protect\citeauthoryear{{Koester}, {Rollenhagen}, {Napiwotzki},
  {Voss}, {Christlieb}, {Homeier}  \& {Reimers}}{{Koester}
  et~al.}{2005}]{2005Koester}
{Koester} D.,  {Rollenhagen} K.,  {Napiwotzki} R.,  {Voss} B.,  {Christlieb}
  N.,  {Homeier} D.,   {Reimers} D.,  2005, \mn@doi [\aap]
  {10.1051/0004-6361:20041927}, \href
  {https://ui.adsabs.harvard.edu/abs/2005A&A...432.1025K} {432, 1025}

\bibitem[\protect\citeauthoryear{{Koester}, {G{\"a}nsicke}  \&
  {Farihi}}{{Koester} et~al.}{2014}]{2014Koester}
{Koester} D.,  {G{\"a}nsicke} B.~T.,   {Farihi} J.,  2014, \mn@doi [\aap]
  {10.1051/0004-6361/201423691}, \href
  {https://ui.adsabs.harvard.edu/abs/2014A&A...566A..34K} {566, A34}

\bibitem[\protect\citeauthoryear{{Koester}, {Kepler}  \& {Irwin}}{{Koester}
  et~al.}{2020}]{2020Koester}
{Koester} D.,  {Kepler} S.~O.,   {Irwin} A.~W.,  2020, \mn@doi [\aap]
  {10.1051/0004-6361/202037530}, \href
  {https://ui.adsabs.harvard.edu/abs/2020A&A...635A.103K} {635, A103}

\bibitem[\protect\citeauthoryear{{Konopacky} et~al.,}{{Konopacky}
  et~al.}{2012}]{2012Konopacky}
{Konopacky} Q.~M.,  et~al., 2012, \mn@doi [\apj] {10.1088/0004-637X/750/1/79},
  \href {https://ui.adsabs.harvard.edu/abs/2012ApJ...750...79K} {750, 79}

\bibitem[\protect\citeauthoryear{{Kumar}, {Rajpurohit}  \&
  {Srivastava}}{{Kumar} et~al.}{2023}]{2023Kumar}
{Kumar} V.,  {Rajpurohit} A.~S.,   {Srivastava} M.~K.,  2023, \mn@doi [arXiv
  e-prints] {10.48550/arXiv.2302.01643}, \href
  {https://ui.adsabs.harvard.edu/abs/2023arXiv230201643K} {p. arXiv:2302.01643}

\bibitem[\protect\citeauthoryear{{Laming}}{{Laming}}{2015}]{2015Laming}
{Laming} J.~M.,  2015, \mn@doi [Living Reviews in Solar Physics]
  {10.1007/lrsp-2015-2}, \href
  {https://ui.adsabs.harvard.edu/abs/2015LRSP...12....2L} {12, 2}

\bibitem[\protect\citeauthoryear{{Lanning}}{{Lanning}}{1982}]{1982Lanning}
{Lanning} H.~H.,  1982, \mn@doi [\apj] {10.1086/159676}, \href
  {https://ui.adsabs.harvard.edu/abs/1982ApJ...253..752L} {253, 752}

\bibitem[\protect\citeauthoryear{{Li}, {Wu}  \& {Wickramasinghe}}{{Li}
  et~al.}{1994}]{1994Li}
{Li} J.~K.,  {Wu} K.~W.,   {Wickramasinghe} D.~T.,  1994, \mn@doi [\mnras]
  {10.1093/mnras/268.1.61}, \href
  {https://ui.adsabs.harvard.edu/abs/1994MNRAS.268...61L} {268, 61}

\bibitem[\protect\citeauthoryear{{Line} et~al.,}{{Line}
  et~al.}{2017}]{2017Line}
{Line} M.~R.,  et~al., 2017, \mn@doi [\apj] {10.3847/1538-4357/aa7ff0}, \href
  {https://ui.adsabs.harvard.edu/abs/2017ApJ...848...83L} {848, 83}

\bibitem[\protect\citeauthoryear{{Littlefair} et~al.,}{{Littlefair}
  et~al.}{2014}]{2014Littlefair}
{Littlefair} S.~P.,  et~al., 2014, \mn@doi [\mnras] {10.1093/mnras/stu1895},
  \href {https://ui.adsabs.harvard.edu/abs/2014MNRAS.445.2106L} {445, 2106}

\bibitem[\protect\citeauthoryear{{Lodders}}{{Lodders}}{1999}]{1999Lodders}
{Lodders} K.,  1999, \mn@doi [\apj] {10.1086/307387}, \href
  {https://ui.adsabs.harvard.edu/abs/1999ApJ...519..793L} {519, 793}

\bibitem[\protect\citeauthoryear{{Lodders}}{{Lodders}}{2002}]{2002Lodders}
{Lodders} K.,  2002, \mn@doi [\apj] {10.1086/342241}, \href
  {https://ui.adsabs.harvard.edu/abs/2002ApJ...577..974L} {577, 974}

\bibitem[\protect\citeauthoryear{{Lodders}}{{Lodders}}{2003}]{2003Lodders}
{Lodders} K.,  2003, \mn@doi [\apj] {10.1086/375492}, \href
  {https://ui.adsabs.harvard.edu/abs/2003ApJ...591.1220L} {591, 1220}

\bibitem[\protect\citeauthoryear{{Longstaff}, {Casewell}, {Wynn}, {Maxted}  \&
  {Helling}}{{Longstaff} et~al.}{2017}]{2017Longstaff}
{Longstaff} E.~S.,  {Casewell} S.~L.,  {Wynn} G.~A.,  {Maxted} P.~F.~L.,
  {Helling} C.,  2017, \mn@doi [\mnras] {10.1093/mnras/stx1786}, \href
  {https://ui.adsabs.harvard.edu/abs/2017MNRAS.471.1728L} {471, 1728}

\bibitem[\protect\citeauthoryear{{Longstaff}, {Casewell}, {Wynn}, {Page},
  {Williams}, {Braker}  \& {Maxted}}{{Longstaff} et~al.}{2019}]{2019Longstaff}
{Longstaff} E.~S.,  {Casewell} S.~L.,  {Wynn} G.~A.,  {Page} K.~L.,  {Williams}
  P.~K.~G.,  {Braker} I.,   {Maxted} P.~F.~L.,  2019, \mn@doi [\mnras]
  {10.1093/mnras/stz127}, \href
  {https://ui.adsabs.harvard.edu/abs/2019MNRAS.484.2566L} {484, 2566}

\bibitem[\protect\citeauthoryear{{Madhusudhan}, {Apai}  \&
  {Gandhi}}{{Madhusudhan} et~al.}{2016}]{2016Madhusudhan}
{Madhusudhan} N.,  {Apai} D.,   {Gandhi} S.,  2016, arXiv e-prints, \href
  {https://ui.adsabs.harvard.edu/abs/2016arXiv161203174M} {p. arXiv:1612.03174}

\bibitem[\protect\citeauthoryear{{Maxted}, {Marsh}, {Moran}, {Dhillon}  \&
  {Hilditch}}{{Maxted} et~al.}{1998}]{1998Maxted}
{Maxted} P.~F.~L.,  {Marsh} T.~R.,  {Moran} C.,  {Dhillon} V.~S.,   {Hilditch}
  R.~W.,  1998, \mn@doi [\mnras] {10.1046/j.1365-8711.1998.02036.x}, \href
  {https://ui.adsabs.harvard.edu/abs/1998MNRAS.300.1225M} {300, 1225}

\bibitem[\protect\citeauthoryear{{Maxted}, {Napiwotzki}, {Dobbie}  \&
  {Burleigh}}{{Maxted} et~al.}{2006}]{2006Maxted}
{Maxted} P.~F.~L.,  {Napiwotzki} R.,  {Dobbie} P.~D.,   {Burleigh} M.~R.,
  2006, \mn@doi [\nat] {10.1038/nature04987}, \href
  {https://ui.adsabs.harvard.edu/abs/2006Natur.442..543M} {442, 543}

\bibitem[\protect\citeauthoryear{{Maxted}, {O'Donoghue}, {Morales-Rueda},
  {Napiwotzki}  \& {Smalley}}{{Maxted} et~al.}{2007}]{2007Maxted}
{Maxted} P.~F.~L.,  {O'Donoghue} D.,  {Morales-Rueda} L.,  {Napiwotzki} R.,
  {Smalley} B.,  2007, \mn@doi [\mnras] {10.1111/j.1365-2966.2007.11564.x},
  \href {https://ui.adsabs.harvard.edu/abs/2007MNRAS.376..919M} {376, 919}

\bibitem[\protect\citeauthoryear{{Mayer} et~al.,}{{Mayer}
  et~al.}{2014}]{2014Mayer}
{Mayer} A.,  et~al., 2014, \mn@doi [\aap] {10.1051/0004-6361/201424465}, \href
  {https://ui.adsabs.harvard.edu/abs/2014A&A...570A.113M} {570, A113}

\bibitem[\protect\citeauthoryear{{Mesquita} \& {Vidotto}}{{Mesquita} \&
  {Vidotto}}{2020}]{2020Mesquita}
{Mesquita} A.~L.,  {Vidotto} A.~A.,  2020, \mn@doi [\mnras]
  {10.1093/mnras/staa798}, \href
  {https://ui.adsabs.harvard.edu/abs/2020MNRAS.494.1297M} {494, 1297}

\bibitem[\protect\citeauthoryear{{Modigliani} et~al.,}{{Modigliani}
  et~al.}{2010}]{2010Modigliani}
{Modigliani} A.,  et~al., 2010, in {Silva} D.~R.,  {Peck} A.~B.,   {Soifer}
  B.~T.,  eds,  Society of Photo-Optical Instrumentation Engineers (SPIE)
  Conference Series Vol. 7737, Observatory Operations: Strategies, Processes,
  and Systems III. p. 773728, \mn@doi{10.1117/12.857211}

\bibitem[\protect\citeauthoryear{{Mohamed} \& {Podsiadlowski}}{{Mohamed} \&
  {Podsiadlowski}}{2007}]{2007Mohamed}
{Mohamed} S.,  {Podsiadlowski} P.,  2007, in {Napiwotzki} R.,  {Burleigh}
  M.~R.,  eds,  Astronomical Society of the Pacific Conference Series Vol. 372,
  15th European Workshop on White Dwarfs. p.~397

\bibitem[\protect\citeauthoryear{{Mohanty} \& {Basri}}{{Mohanty} \&
  {Basri}}{2003}]{2003Mohanty}
{Mohanty} S.,  {Basri} G.,  2003, \mn@doi [\apj] {10.1086/345097}, \href
  {https://ui.adsabs.harvard.edu/abs/2003ApJ...583..451M} {583, 451}

\bibitem[\protect\citeauthoryear{{Mohanty}, {Basri}, {Shu}, {Allard}  \&
  {Chabrier}}{{Mohanty} et~al.}{2002}]{2002Mohanty}
{Mohanty} S.,  {Basri} G.,  {Shu} F.,  {Allard} F.,   {Chabrier} G.,  2002,
  \mn@doi [\apj] {10.1086/339911}, \href
  {https://ui.adsabs.harvard.edu/abs/2002ApJ...571..469M} {571, 469}

\bibitem[\protect\citeauthoryear{{Morin}, {Dormy}, {Schrinner}  \&
  {Donati}}{{Morin} et~al.}{2011}]{2011Morin}
{Morin} J.,  {Dormy} E.,  {Schrinner} M.,   {Donati} J.~F.,  2011, \mn@doi
  [\mnras] {10.1111/j.1745-3933.2011.01159.x}, \href
  {https://ui.adsabs.harvard.edu/abs/2011MNRAS.418L.133M} {418, L133}

\bibitem[\protect\citeauthoryear{{Napiwotzki} et~al.,}{{Napiwotzki}
  et~al.}{2001}]{2001Napiwotzki}
{Napiwotzki} R.,  et~al., 2001, \mn@doi [Astronomische Nachrichten]
  {10.1002/1521-3994(200112)322:5/6\textless{}411::AID-ASNA411\textgreater{}3.0.CO;2-I},
  \href {https://ui.adsabs.harvard.edu/abs/2001AN....322..411N} {322, 411}

\bibitem[\protect\citeauthoryear{{Napiwotzki} et~al.,}{{Napiwotzki}
  et~al.}{2003}]{2003Napiwotzki}
{Napiwotzki} R.,  et~al., 2003, The Messenger, \href
  {https://ui.adsabs.harvard.edu/abs/2003Msngr.112...25N} {112, 25}

\bibitem[\protect\citeauthoryear{{Newville}, {Stensitzki}, {Allen}, {Rawlik},
  {Ingargiola}  \& {Nelson}}{{Newville} et~al.}{2016}]{2016Newville}
{Newville} M.,  {Stensitzki} T.,  {Allen} D.~B.,  {Rawlik} M.,  {Ingargiola}
  A.,   {Nelson} A.,  2016, {Lmfit: Non-Linear Least-Square Minimization and
  Curve-Fitting for Python} (\mn@eprint {ascl} {1606.014})

\bibitem[\protect\citeauthoryear{{Nordstr{\"o}m} et~al.,}{{Nordstr{\"o}m}
  et~al.}{2004}]{2004Nordstrom}
{Nordstr{\"o}m} B.,  et~al., 2004, \mn@doi [\aap] {10.1051/0004-6361:20035959},
  \href {https://ui.adsabs.harvard.edu/abs/2004A&A...418..989N} {418, 989}

\bibitem[\protect\citeauthoryear{{Oomen}, {Van Winckel}, {Pols}  \&
  {Nelemans}}{{Oomen} et~al.}{2019}]{2019Oomen}
{Oomen} G.-M.,  {Van Winckel} H.,  {Pols} O.,   {Nelemans} G.,  2019, \mn@doi
  [\aap] {10.1051/0004-6361/201935853}, \href
  {https://ui.adsabs.harvard.edu/abs/2019A&A...629A..49O} {629, A49}

\bibitem[\protect\citeauthoryear{{Oomen}, {Pols}, {Van Winckel}  \&
  {Nelemans}}{{Oomen} et~al.}{2020}]{2020Oomen}
{Oomen} G.-M.,  {Pols} O.,  {Van Winckel} H.,   {Nelemans} G.,  2020, \mn@doi
  [\aap] {10.1051/0004-6361/202038341}, \href
  {https://ui.adsabs.harvard.edu/abs/2020A&A...642A.234O} {642, A234}

\bibitem[\protect\citeauthoryear{{Owen} \& {Alvarez}}{{Owen} \&
  {Alvarez}}{2016}]{2016Owen_a}
{Owen} J.~E.,  {Alvarez} M.~A.,  2016, \mn@doi [\apj]
  {10.3847/0004-637X/816/1/34}, \href
  {https://ui.adsabs.harvard.edu/abs/2016ApJ...816...34O} {816, 34}

\bibitem[\protect\citeauthoryear{{Owen} \& {Wu}}{{Owen} \&
  {Wu}}{2016}]{2016Owen_b}
{Owen} J.~E.,  {Wu} Y.,  2016, \mn@doi [\apj] {10.3847/0004-637X/817/2/107},
  \href {https://ui.adsabs.harvard.edu/abs/2016ApJ...817..107O} {817, 107}

\bibitem[\protect\citeauthoryear{{Parsons} et~al.,}{{Parsons}
  et~al.}{2017a}]{2017Parsons_b}
{Parsons} S.~G.,  et~al., 2017a, \mn@doi [\mnras] {10.1093/mnras/stx1522},
  \href {https://ui.adsabs.harvard.edu/abs/2017MNRAS.470.4473P} {470, 4473}

\bibitem[\protect\citeauthoryear{{Parsons} et~al.,}{{Parsons}
  et~al.}{2017b}]{2017Parsons_a}
{Parsons} S.~G.,  et~al., 2017b, \mn@doi [\mnras] {10.1093/mnras/stx1610},
  \href {https://ui.adsabs.harvard.edu/abs/2017MNRAS.471..976P} {471, 976}

\bibitem[\protect\citeauthoryear{{Parsons} et~al.,}{{Parsons}
  et~al.}{2018}]{2018Parsons}
{Parsons} S.~G.,  et~al., 2018, \mn@doi [\mnras] {10.1093/mnras/sty2345}, \href
  {https://ui.adsabs.harvard.edu/abs/2018MNRAS.481.1083P} {481, 1083}

\bibitem[\protect\citeauthoryear{{Parsons}, {G{\"a}nsicke}, {Schreiber},
  {Marsh}, {Ashley}, {Breedt}, {Littlefair}  \& {Meusinger}}{{Parsons}
  et~al.}{2021}]{2021Parsons}
{Parsons} S.~G.,  {G{\"a}nsicke} B.~T.,  {Schreiber} M.~R.,  {Marsh} T.~R.,
  {Ashley} R.~P.,  {Breedt} E.,  {Littlefair} S.~P.,   {Meusinger} H.,  2021,
  \mn@doi [\mnras] {10.1093/mnras/stab284}, \href
  {https://ui.adsabs.harvard.edu/abs/2021MNRAS.502.4305P} {502, 4305}

\bibitem[\protect\citeauthoryear{{Pecaut} \& {Mamajek}}{{Pecaut} \&
  {Mamajek}}{2013}]{2013Pecaut}
{Pecaut} M.~J.,  {Mamajek} E.~E.,  2013, \mn@doi [\apjs]
  {10.1088/0067-0049/208/1/9}, \href
  {https://ui.adsabs.harvard.edu/abs/2013ApJS..208....9P} {208, 9}

\bibitem[\protect\citeauthoryear{{Pineda}, {Hallinan}, {Kirkpatrick}, {Cotter},
  {Kao}  \& {Mooley}}{{Pineda} et~al.}{2016}]{Pineda2016}
{Pineda} J.~S.,  {Hallinan} G.,  {Kirkpatrick} J.~D.,  {Cotter} G.,  {Kao}
  M.~M.,   {Mooley} K.,  2016, \mn@doi [\apj] {10.3847/0004-637X/826/1/73},
  \href {https://ui.adsabs.harvard.edu/abs/2016ApJ...826...73P} {826, 73}

\bibitem[\protect\citeauthoryear{{Pineda}, {Hallinan}  \& {Kao}}{{Pineda}
  et~al.}{2017}]{2017Pineda}
{Pineda} J.~S.,  {Hallinan} G.,   {Kao} M.~M.,  2017, \mn@doi [\apj]
  {10.3847/1538-4357/aa8596}, \href
  {https://ui.adsabs.harvard.edu/abs/2017ApJ...846...75P} {846, 75}

\bibitem[\protect\citeauthoryear{{Reiners}}{{Reiners}}{2012}]{2012Reiners_b}
{Reiners} A.,  2012, \mn@doi [Living Reviews in Solar Physics]
  {10.12942/lrsp-2012-1}, \href
  {https://ui.adsabs.harvard.edu/abs/2012LRSP....9....1R} {9, 1}

\bibitem[\protect\citeauthoryear{{Reiners} \& {Basri}}{{Reiners} \&
  {Basri}}{2008}]{2008Reiners}
{Reiners} A.,  {Basri} G.,  2008, \mn@doi [\apj] {10.1086/590073}, \href
  {https://ui.adsabs.harvard.edu/abs/2008ApJ...684.1390R} {684, 1390}

\bibitem[\protect\citeauthoryear{{Reiners} \& {Basri}}{{Reiners} \&
  {Basri}}{2010}]{2010Reiners}
{Reiners} A.,  {Basri} G.,  2010, \mn@doi [\apj] {10.1088/0004-637X/710/2/924},
  \href {https://ui.adsabs.harvard.edu/abs/2010ApJ...710..924R} {710, 924}

\bibitem[\protect\citeauthoryear{{Reiners} \& {Mohanty}}{{Reiners} \&
  {Mohanty}}{2012}]{2012Reiners}
{Reiners} A.,  {Mohanty} S.,  2012, \mn@doi [\apj]
  {10.1088/0004-637X/746/1/43}, \href
  {https://ui.adsabs.harvard.edu/abs/2012ApJ...746...43R} {746, 43}

\bibitem[\protect\citeauthoryear{{Reiners}, {Sch{\"u}ssler}  \&
  {Passegger}}{{Reiners} et~al.}{2014}]{2014Reiners}
{Reiners} A.,  {Sch{\"u}ssler} M.,   {Passegger} V.~M.,  2014, \mn@doi [\apj]
  {10.1088/0004-637X/794/2/144}, \href
  {https://ui.adsabs.harvard.edu/abs/2014ApJ...794..144R} {794, 144}

\bibitem[\protect\citeauthoryear{{Reiners} et~al.,}{{Reiners}
  et~al.}{2022}]{2022Reiners}
{Reiners} A.,  et~al., 2022, arXiv e-prints, \href
  {https://ui.adsabs.harvard.edu/abs/2022arXiv220400342R} {p. arXiv:2204.00342}

\bibitem[\protect\citeauthoryear{{Ribeiro}, {Baptista}, {Kafka}, {Dufour},
  {Gianninas}  \& {Fontaine}}{{Ribeiro} et~al.}{2013}]{2013Ribeiro}
{Ribeiro} T.,  {Baptista} R.,  {Kafka} S.,  {Dufour} P.,  {Gianninas} A.,
  {Fontaine} G.,  2013, \mn@doi [\aap] {10.1051/0004-6361/201220340}, \href
  {https://ui.adsabs.harvard.edu/abs/2013A&A...556A..34R} {556, A34}

\bibitem[\protect\citeauthoryear{{Ricker} et~al.,}{{Ricker}
  et~al.}{2015}]{2015Ricker}
{Ricker} G.~R.,  et~al., 2015, \mn@doi [Journal of Astronomical Telescopes,
  Instruments, and Systems] {10.1117/1.JATIS.1.1.014003}, \href
  {https://ui.adsabs.harvard.edu/abs/2015JATIS...1a4003R} {1, 014003}

\bibitem[\protect\citeauthoryear{{Ridgway} et~al.,}{{Ridgway}
  et~al.}{2023}]{2023Ridgway}
{Ridgway} R.~J.,  et~al., 2023, \mn@doi [\mnras] {10.1093/mnras/stac3105},
  \href {https://ui.adsabs.harvard.edu/abs/2023MNRAS.518.2472R} {518, 2472}

\bibitem[\protect\citeauthoryear{{Route} \& {Wolszczan}}{{Route} \&
  {Wolszczan}}{2012}]{2012Route}
{Route} M.,  {Wolszczan} A.,  2012, \mn@doi [\apjl]
  {10.1088/2041-8205/747/2/L22}, \href
  {https://ui.adsabs.harvard.edu/abs/2012ApJ...747L..22R} {747, L22}

\bibitem[\protect\citeauthoryear{{Saffer}, {Wade}, {Liebert}, {Green}, {Sion},
  {Bechtold}, {Foss}  \& {Kidder}}{{Saffer} et~al.}{1993}]{1993Saffer}
{Saffer} R.~A.,  {Wade} R.~A.,  {Liebert} J.,  {Green} R.~F.,  {Sion} E.~M.,
  {Bechtold} J.,  {Foss} D.,   {Kidder} K.,  1993, \mn@doi [\aj]
  {10.1086/116569}, \href
  {https://ui.adsabs.harvard.edu/abs/1993AJ....105.1945S} {105, 1945}

\bibitem[\protect\citeauthoryear{{Schmidt} et~al.,}{{Schmidt}
  et~al.}{2005}]{2005Schmidt}
{Schmidt} G.~D.,  et~al., 2005, \mn@doi [\apj] {10.1086/431969}, \href
  {https://ui.adsabs.harvard.edu/abs/2005ApJ...630.1037S} {630, 1037}

\bibitem[\protect\citeauthoryear{{Schmidt}, {Cruz}, {Bongiorno}, {Liebert}  \&
  {Reid}}{{Schmidt} et~al.}{2007a}]{2007Schmidt_a}
{Schmidt} S.~J.,  {Cruz} K.~L.,  {Bongiorno} B.~J.,  {Liebert} J.,   {Reid}
  I.~N.,  2007a, \mn@doi [\aj] {10.1086/512158}, \href
  {https://ui.adsabs.harvard.edu/abs/2007AJ....133.2258S} {133, 2258}

\bibitem[\protect\citeauthoryear{{Schmidt}, {Szkody}, {Henden}, {Anderson},
  {Lamb}, {Margon}  \& {Schneider}}{{Schmidt} et~al.}{2007b}]{2007Schmidt_b}
{Schmidt} G.~D.,  {Szkody} P.,  {Henden} A.,  {Anderson} S.~F.,  {Lamb} D.~Q.,
  {Margon} B.,   {Schneider} D.~P.,  2007b, \mn@doi [\apj] {10.1086/509613},
  \href {https://ui.adsabs.harvard.edu/abs/2007ApJ...654..521S} {654, 521}

\bibitem[\protect\citeauthoryear{{Schmidt}, {Hawley}, {West}, {Bochanski},
  {Davenport}, {Ge}  \& {Schneider}}{{Schmidt} et~al.}{2015}]{2015Schmidt}
{Schmidt} S.~J.,  {Hawley} S.~L.,  {West} A.~A.,  {Bochanski} J.~J.,
  {Davenport} J. R.~A.,  {Ge} J.,   {Schneider} D.~P.,  2015, \mn@doi [\aj]
  {10.1088/0004-6256/149/5/158}, \href
  {https://ui.adsabs.harvard.edu/abs/2015AJ....149..158S} {149, 158}

\bibitem[\protect\citeauthoryear{{Schmidt} et~al.,}{{Schmidt}
  et~al.}{2016}]{2016Schmidt}
{Schmidt} S.~J.,  et~al., 2016, \mn@doi [\apjl] {10.3847/2041-8205/828/2/L22},
  \href {https://ui.adsabs.harvard.edu/abs/2016ApJ...828L..22S} {828, L22}

\bibitem[\protect\citeauthoryear{{Scholz}, {Kostov}, {Jayawardhana}  \&
  {Mu{\v{z}}i{\'c}}}{{Scholz} et~al.}{2015}]{2015Scholz}
{Scholz} A.,  {Kostov} V.,  {Jayawardhana} R.,   {Mu{\v{z}}i{\'c}} K.,  2015,
  \mn@doi [\apjl] {10.1088/2041-8205/809/2/L29}, \href
  {https://ui.adsabs.harvard.edu/abs/2015ApJ...809L..29S} {809, L29}

\bibitem[\protect\citeauthoryear{{Schrijver}}{{Schrijver}}{2009}]{2009Schrijver}
{Schrijver} C.~J.,  2009, \mn@doi [\apjl] {10.1088/0004-637X/699/2/L148}, \href
  {https://ui.adsabs.harvard.edu/abs/2009ApJ...699L.148S} {699, L148}

\bibitem[\protect\citeauthoryear{{Schwope}, {Brunner}, {Hambaryan}  \&
  {Schwarz}}{{Schwope} et~al.}{2002}]{2002Schwope}
{Schwope} A.~D.,  {Brunner} H.,  {Hambaryan} V.,   {Schwarz} R.,  2002, in
  {G{\"a}nsicke} B.~T.,  {Beuermann} K.,   {Reinsch} K.,  eds,  Astronomical
  Society of the Pacific Conference Series Vol. 261, The Physics of Cataclysmic
  Variables and Related Objects. p.~102

\bibitem[\protect\citeauthoryear{{Schwope}, {Nebot Gomez-Moran}, {Schreiber}
  \& {G{\"a}nsicke}}{{Schwope} et~al.}{2009}]{2009Schwope}
{Schwope} A.~D.,  {Nebot Gomez-Moran} A.,  {Schreiber} M.~R.,   {G{\"a}nsicke}
  B.~T.,  2009, \mn@doi [\aap] {10.1051/0004-6361/200911699}, \href
  {https://ui.adsabs.harvard.edu/abs/2009A&A...500..867S} {500, 867}

\bibitem[\protect\citeauthoryear{{Skumanich}}{{Skumanich}}{1972}]{1972Skumanich}
{Skumanich} A.,  1972, \mn@doi [\apj] {10.1086/151310}, \href
  {https://ui.adsabs.harvard.edu/abs/1972ApJ...171..565S} {171, 565}

\bibitem[\protect\citeauthoryear{{Steele} et~al.,}{{Steele}
  et~al.}{2013}]{2013Steele}
{Steele} P.~R.,  et~al., 2013, \mn@doi [\mnras] {10.1093/mnras/sts620}, \href
  {https://ui.adsabs.harvard.edu/abs/2013MNRAS.429.3492S} {429, 3492}

\bibitem[\protect\citeauthoryear{{Stelzer}, {Micela}, {Flaccomio},
  {Neuh{\"a}user}  \& {Jayawardhana}}{{Stelzer} et~al.}{2006}]{2006Stelzer}
{Stelzer} B.,  {Micela} G.,  {Flaccomio} E.,  {Neuh{\"a}user} R.,
  {Jayawardhana} R.,  2006, \mn@doi [\aap] {10.1051/0004-6361:20053677}, \href
  {https://ui.adsabs.harvard.edu/abs/2006A&A...448..293S} {448, 293}

\bibitem[\protect\citeauthoryear{{Stelzer} et~al.,}{{Stelzer}
  et~al.}{2012}]{2012Stelzer}
{Stelzer} B.,  et~al., 2012, \mn@doi [\aap] {10.1051/0004-6361/201118097},
  \href {https://ui.adsabs.harvard.edu/abs/2012A&A...537A..94S} {537, A94}

\bibitem[\protect\citeauthoryear{{Stelzer}, {de Martino}, {Casewell}, {Wynn}
  \& {Roy}}{{Stelzer} et~al.}{2017}]{2017Stelzer}
{Stelzer} B.,  {de Martino} D.,  {Casewell} S.~L.,  {Wynn} G.~A.,   {Roy} M.,
  2017, \mn@doi [\aap] {10.1051/0004-6361/201630038}, \href
  {https://ui.adsabs.harvard.edu/abs/2017A&A...598L...6S} {598, L6}

\bibitem[\protect\citeauthoryear{{Swan}, {Farihi}, {Koester}, {Hollands},
  {Parsons}, {Cauley}, {Redfield}  \& {G{\"a}nsicke}}{{Swan}
  et~al.}{2019}]{2019Swan}
{Swan} A.,  {Farihi} J.,  {Koester} D.,  {Hollands} M.,  {Parsons} S.,
  {Cauley} P.~W.,  {Redfield} S.,   {G{\"a}nsicke} B.~T.,  2019, \mn@doi
  [\mnras] {10.1093/mnras/stz2337}, \href
  {https://ui.adsabs.harvard.edu/abs/2019MNRAS.490..202S} {490, 202}

\bibitem[\protect\citeauthoryear{{Tannock} et~al.,}{{Tannock}
  et~al.}{2021}]{2021Tannock}
{Tannock} M.~E.,  et~al., 2021, \mn@doi [\aj] {10.3847/1538-3881/abeb67}, \href
  {https://ui.adsabs.harvard.edu/abs/2021AJ....161..224T} {161, 224}

\bibitem[\protect\citeauthoryear{{Tappert}, {G{\"a}nsicke},
  {Rebassa-Mansergas}, {Schmidtobreick}  \& {Schreiber}}{{Tappert}
  et~al.}{2011a}]{2011Tappert_b}
{Tappert} C.,  {G{\"a}nsicke} B.~T.,  {Rebassa-Mansergas} A.,  {Schmidtobreick}
  L.,   {Schreiber} M.~R.,  2011a, \mn@doi [\aap]
  {10.1051/0004-6361/201116833}, \href
  {https://ui.adsabs.harvard.edu/abs/2011A&A...531A.113T} {531, A113}

\bibitem[\protect\citeauthoryear{{Tappert}, {G{\"a}nsicke}, {Schmidtobreick}
  \& {Ribeiro}}{{Tappert} et~al.}{2011b}]{2011Tappert_a}
{Tappert} C.,  {G{\"a}nsicke} B.~T.,  {Schmidtobreick} L.,   {Ribeiro} T.,
  2011b, \mn@doi [\aap] {10.1051/0004-6361/201116436}, \href
  {https://ui.adsabs.harvard.edu/abs/2011A&A...532A.129T} {532, A129}

\bibitem[\protect\citeauthoryear{{Tremblay} \& {Bergeron}}{{Tremblay} \&
  {Bergeron}}{2009}]{2009Tremblay}
{Tremblay} P.~E.,  {Bergeron} P.,  2009, \mn@doi [\apj]
  {10.1088/0004-637X/696/2/1755}, \href
  {https://ui.adsabs.harvard.edu/abs/2009ApJ...696.1755T} {696, 1755}

\bibitem[\protect\citeauthoryear{{Tsuboi}, {Maeda}, {Feigelson}, {Garmire},
  {Chartas}, {Mori}  \& {Pravdo}}{{Tsuboi} et~al.}{2003}]{2003Tsuboi}
{Tsuboi} Y.,  {Maeda} Y.,  {Feigelson} E.~D.,  {Garmire} G.~P.,  {Chartas} G.,
  {Mori} K.,   {Pravdo} S.~H.,  2003, \mn@doi [\apjl] {10.1086/375017}, \href
  {https://ui.adsabs.harvard.edu/abs/2003ApJ...587L..51T} {587, L51}

\bibitem[\protect\citeauthoryear{{Van Winckel}, {Mathis}  \& {Waelkens}}{{Van
  Winckel} et~al.}{1992}]{1992vanWinckel}
{Van Winckel} H.,  {Mathis} J.~S.,   {Waelkens} C.,  1992, \nat, \href
  {https://ui.adsabs.harvard.edu/abs/1992Natur.356..500V} {356, 500}

\bibitem[\protect\citeauthoryear{{Van Winckel}, {Waelkens}, {Waters},
  {Molster}, {Udry}  \& {Bakker}}{{Van Winckel} et~al.}{1998}]{1998vanWinckel}
{Van Winckel} H.,  {Waelkens} C.,  {Waters} L. B.~F.~M.,  {Molster} F.~J.,
  {Udry} S.,   {Bakker} E.~J.,  1998, \aap, \href
  {https://ui.adsabs.harvard.edu/abs/1998A&A...336L..17V} {336, L17}

\bibitem[\protect\citeauthoryear{{Vauclair}, {Vauclair}  \&
  {Greenstein}}{{Vauclair} et~al.}{1979}]{1979Vauclair}
{Vauclair} G.,  {Vauclair} S.,   {Greenstein} J.~L.,  1979, \aap, \href
  {https://ui.adsabs.harvard.edu/abs/1979A&A....80...79V} {80, 79}

\bibitem[\protect\citeauthoryear{{Vernet} et~al.,}{{Vernet}
  et~al.}{2011}]{2011Vernet}
{Vernet} J.,  et~al., 2011, \mn@doi [\aap] {10.1051/0004-6361/201117752}, \href
  {https://ui.adsabs.harvard.edu/abs/2011A&A...536A.105V} {536, A105}

\bibitem[\protect\citeauthoryear{{Vidotto}}{{Vidotto}}{2021}]{2021Vidotto}
{Vidotto} A.~A.,  2021, \mn@doi [Living Reviews in Solar Physics]
  {10.1007/s41116-021-00029-w}, \href
  {https://ui.adsabs.harvard.edu/abs/2021LRSP...18....3V} {18, 3}

\bibitem[\protect\citeauthoryear{{Vidotto}, {Jardine}, {Morin}, {Donati},
  {Opher}  \& {Gombosi}}{{Vidotto} et~al.}{2014}]{2014Vidotto}
{Vidotto} A.~A.,  {Jardine} M.,  {Morin} J.,  {Donati} J.~F.,  {Opher} M.,
  {Gombosi} T.~I.,  2014, \mn@doi [\mnras] {10.1093/mnras/stt2265}, \href
  {https://ui.adsabs.harvard.edu/abs/2014MNRAS.438.1162V} {438, 1162}

\bibitem[\protect\citeauthoryear{{Waelkens}, {Van Winckel}, {Waters}  \&
  {Bakker}}{{Waelkens} et~al.}{1996}]{1996Waelkens}
{Waelkens} C.,  {Van Winckel} H.,  {Waters} L.~B.~F.~M.,   {Bakker} E.~J.,
  1996, \mn@doi [\aap] {10.48550/arXiv.astro-ph/9609058}, \href
  {https://ui.adsabs.harvard.edu/abs/1996A&A...314L..17W} {314, L17}

\bibitem[\protect\citeauthoryear{{Walters}, {Farihi}, {Marsh}, {Breedt},
  {Cauley}, {von Hippel}  \& {Hermes}}{{Walters} et~al.}{2022}]{2022Walters}
{Walters} N.,  {Farihi} J.,  {Marsh} T.~R.,  {Breedt} E.,  {Cauley} P.~W.,
  {von Hippel} T.,   {Hermes} J.~J.,  2022, arXiv e-prints, \href
  {https://ui.adsabs.harvard.edu/abs/2022arXiv220707022W} {p. arXiv:2207.07022}

\bibitem[\protect\citeauthoryear{{Waters}, {Trams}  \& {Waelkens}}{{Waters}
  et~al.}{1992}]{1992Waters}
{Waters} L.~B.~F.~M.,  {Trams} N.~R.,   {Waelkens} C.,  1992, \aap, \href
  {https://ui.adsabs.harvard.edu/abs/1992A&A...262L..37W} {262, L37}

\bibitem[\protect\citeauthoryear{{Webbink} \& {Wickramasinghe}}{{Webbink} \&
  {Wickramasinghe}}{2002}]{2002Webbink}
{Webbink} R.~F.,  {Wickramasinghe} D.~T.,  2002, \mn@doi [\mnras]
  {10.1046/j.1365-8711.2002.05495.x}, \href
  {https://ui.adsabs.harvard.edu/abs/2002MNRAS.335....1W} {335, 1}

\bibitem[\protect\citeauthoryear{{Webbink} \& {Wickramasinghe}}{{Webbink} \&
  {Wickramasinghe}}{2005}]{2005Webbink}
{Webbink} R.~F.,  {Wickramasinghe} D.~T.,  2005, in {Hameury} J.~M.,  {Lasota}
  J.~P.,  eds,  Astronomical Society of the Pacific Conference Series Vol. 330,
  The Astrophysics of Cataclysmic Variables and Related Objects. p.~137

\bibitem[\protect\citeauthoryear{{Williams}, {Cook}  \& {Berger}}{{Williams}
  et~al.}{2014}]{2014Williams}
{Williams} P.~K.~G.,  {Cook} B.~A.,   {Berger} E.,  2014, \mn@doi [\apj]
  {10.1088/0004-637X/785/1/9}, \href
  {https://ui.adsabs.harvard.edu/abs/2014ApJ...785....9W} {785, 9}

\bibitem[\protect\citeauthoryear{{Wood}, {M{\"u}ller}, {Zank}  \&
  {Linsky}}{{Wood} et~al.}{2002}]{2002Wood}
{Wood} B.~E.,  {M{\"u}ller} H.-R.,  {Zank} G.~P.,   {Linsky} J.~L.,  2002,
  \mn@doi [\apj] {10.1086/340797}, \href
  {https://ui.adsabs.harvard.edu/abs/2002ApJ...574..412W} {574, 412}

\bibitem[\protect\citeauthoryear{{Wood}, {M{\"u}ller}, {Zank}, {Linsky}  \&
  {Redfield}}{{Wood} et~al.}{2005}]{2005Wood}
{Wood} B.~E.,  {M{\"u}ller} H.~R.,  {Zank} G.~P.,  {Linsky} J.~L.,   {Redfield}
  S.,  2005, \mn@doi [\apjl] {10.1086/432716}, \href
  {https://ui.adsabs.harvard.edu/abs/2005ApJ...628L.143W} {628, L143}

\bibitem[\protect\citeauthoryear{{Wood}, {Laming}, {Warren}  \&
  {Poppenhaeger}}{{Wood} et~al.}{2018}]{2018Wood}
{Wood} B.~E.,  {Laming} J.~M.,  {Warren} H.~P.,   {Poppenhaeger} K.,  2018,
  \mn@doi [\apj] {10.3847/1538-4357/aaccf6}, \href
  {https://ui.adsabs.harvard.edu/abs/2018ApJ...862...66W} {862, 66}

\bibitem[\protect\citeauthoryear{{Wood} et~al.,}{{Wood}
  et~al.}{2021}]{2021Wood}
{Wood} B.~E.,  et~al., 2021, \mn@doi [\apj] {10.3847/1538-4357/abfda5}, \href
  {https://ui.adsabs.harvard.edu/abs/2021ApJ...915...37W} {915, 37}

\bibitem[\protect\citeauthoryear{{Wright}, {Drake}, {Mamajek}  \&
  {Henry}}{{Wright} et~al.}{2011}]{2011Wright}
{Wright} N.~J.,  {Drake} J.~J.,  {Mamajek} E.~E.,   {Henry} G.~W.,  2011,
  \mn@doi [\apj] {10.1088/0004-637X/743/1/48}, \href
  {https://ui.adsabs.harvard.edu/abs/2011ApJ...743...48W} {743, 48}

\bibitem[\protect\citeauthoryear{{Wright}, {Newton}, {Williams}, {Drake}  \&
  {Yadav}}{{Wright} et~al.}{2018}]{2018Wright}
{Wright} N.~J.,  {Newton} E.~R.,  {Williams} P. K.~G.,  {Drake} J.~J.,
  {Yadav} R.~K.,  2018, \mn@doi [\mnras] {10.1093/mnras/sty1670}, \href
  {https://ui.adsabs.harvard.edu/abs/2018MNRAS.479.2351W} {479, 2351}

\bibitem[\protect\citeauthoryear{{Zijlstra}, {Chapman}, {te Lintel Hekkert},
  {Likkel}, {Comeron}, {Norris}, {Molster}  \& {Cohen}}{{Zijlstra}
  et~al.}{2001}]{2001Zijlstra}
{Zijlstra} A.~A.,  {Chapman} J.~M.,  {te Lintel Hekkert} P.,  {Likkel} L.,
  {Comeron} F.,  {Norris} R.~P.,  {Molster} F.~J.,   {Cohen} R.~J.,  2001,
  \mn@doi [\mnras] {10.1046/j.1365-8711.2001.04113.x}, \href
  {https://ui.adsabs.harvard.edu/abs/2001MNRAS.322..280Z} {322, 280}

\bibitem[\protect\citeauthoryear{{Zuckerman}, {Koester}, {Reid}  \&
  {H{\"u}nsch}}{{Zuckerman} et~al.}{2003}]{2003Zuckerman}
{Zuckerman} B.,  {Koester} D.,  {Reid} I.~N.,   {H{\"u}nsch} M.,  2003, \mn@doi
  [\apj] {10.1086/377492}, \href
  {https://ui.adsabs.harvard.edu/abs/2003ApJ...596..477Z} {596, 477}

\bibitem[\protect\citeauthoryear{{Zuckerman}, {Koester}, {Melis}, {Hansen}  \&
  {Jura}}{{Zuckerman} et~al.}{2007}]{2007Zuckerman}
{Zuckerman} B.,  {Koester} D.,  {Melis} C.,  {Hansen} B.~M.,   {Jura} M.,
  2007, \mn@doi [\apj] {10.1086/522223}, \href
  {https://ui.adsabs.harvard.edu/abs/2007ApJ...671..872Z} {671, 872}

\bibitem[\protect\citeauthoryear{{de Val-Borro}, {Karovska}  \& {Sasselov}}{{de
  Val-Borro} et~al.}{2009}]{2009deValBorro}
{de Val-Borro} M.,  {Karovska} M.,   {Sasselov} D.,  2009, \mn@doi [\apj]
  {10.1088/0004-637X/700/2/1148}, \href
  {https://ui.adsabs.harvard.edu/abs/2009ApJ...700.1148D} {700, 1148}

\bibitem[\protect\citeauthoryear{{van Vledder}, {van der Vlugt}, {Holwerda},
  {Kenworthy}, {Bouwens}  \& {Trenti}}{{van Vledder}
  et~al.}{2016}]{2016vanVledder}
{van Vledder} I.,  {van der Vlugt} D.,  {Holwerda} B.~W.,  {Kenworthy} M.~A.,
  {Bouwens} R.~J.,   {Trenti} M.,  2016, \mn@doi [\mnras]
  {10.1093/mnras/stw258}, \href
  {https://ui.adsabs.harvard.edu/abs/2016MNRAS.458..425V} {458, 425}

\makeatother
\end{thebibliography}

\bsp    % typesetting comment
\label{lastpage}
\end{document}